\documentclass{aa}  

\usepackage{graphicx}
\usepackage{txfonts}
\usepackage{CJK}

\usepackage{enumitem}
\usepackage{amsmath}
\usepackage{amssymb}
\usepackage{natbib}
\usepackage{hyperref}
\hypersetup{
    colorlinks=true,
    linkcolor=blue,
    filecolor=blue,      
    urlcolor=blue,
    citecolor=blue
}

\newcommand   {\EBV}     {E(B\,{-}\,V)}
\newcommand   {\KI}      {\rm K\,\uppercase\expandafter{\romannumeral1}}
\newcommand   {\NI}      {\rm N\,\uppercase\expandafter{\romannumeral1}}
\newcommand   {\FeI}     {\rm Fe\,\uppercase\expandafter{\romannumeral1}}
\newcommand   {\TiI}     {\rm Ti\,\uppercase\expandafter{\romannumeral1}}
\newcommand   {\CaI}     {\rm Ca\,\uppercase\expandafter{\romannumeral1}}
\newcommand   {\EJKs}    {E(J\,{-}\,K_{\rm S})}
\newcommand   {\AKs}     {A_{\rm K_S}}
\newcommand   {\Av}      {A_{\rm V}}
\newcommand   {\Rv}      {R_{\rm V}}
\newcommand   {\Ks}      {K_{\rm S}}
\newcommand   {\Gaia}    {\it Gaia}
\newcommand   {\Teff}    {T_{\rm eff}}
\newcommand   {\logg}    {{\rm log}\,g}

\newcommand   {\xp}      {x^{\prime}}
\newcommand   {\kms}     {\rm km\,s^{-1}}

\newcommand   {\sspect}  {\sigma_{\rm spect}}
\newcommand   {\snoise}  {\sigma_{\rm noise}}
\newcommand   {\sew}     {\sigma_{\rm EW}}

\begin{document}

\begin{CJK*}{UTF8}{gbsn}

\title{The diffuse interstellar band around 8620 {\AA}}
   
\subtitle{I. Methods and application to the GIBS data set}

\author{H. Zhao (赵赫)\inst{1} 
        \and
        M. Schultheis\inst{1}
        \and
        A. Recio-Blanco\inst{1}
        \and
        G. Kordopatis\inst{1}
        \and 
        P. de Laverny\inst{1}
        \and
        A. Rojas-Arriagada\inst{2,3}
        \and
        M. Zoccali \inst{2,3}
        \and
        F. Surot \inst{4}
        \and
        E. Valenti \inst{5,6}
        }
\institute{University C\^ote d'Azur, Observatory of the C\^ote d'Azur, CNRS, Lagrange
          Laboratory, Observatory Bd, CS 34229, \\
          06304 Nice cedex 4, France \\
          \email{he.zhao@oca.eu,mathias.schultheis@oca.eu}
          \and
          Instituto de Astrof\'isica, Facultad de F\'isica, Pontificia, Universidad Cat\'olica de Chile, Av. Vicu\~na Mackenna 4860, Santiago,  Chile
          \and
          Millennium Institute of Astrophysics, Av. Vicu\~na Mackenna 4860, 782-0436 Macul, Santiago, Chile
          \and
          Departamento de Astrof\'{i}sica, Universidad de La Laguna, E\--38205, La Laguna Tenerife, Spain
          \and
          European Southern Observatory, Karl Schwarschild-Stra{\ss}e 2, D-85748 Garching bei M\"unchen, Germany
          \and
          Excellence Cluster ORIGINS, Boltzmann\--Stra\ss e 2, D\--85748 Garching bei M\"{u}nchen, Germany
      }

\date{Received; accepted for A\&A}
 
\abstract
{Diffuse interstellar bands (DIBs) are interstellar absorption features that widely exist in the optical and near-infrared 
wavelength range. DIBs play an important role in the lifecycle of the interstellar medium and can also be used to trace the 
Galactic structure.}
{We developed a set of procedures to automatically detect and measure the DIB around 8620\,{\AA} (the {\it Gaia DIB}) for a wide range of temperatures. 
The method was tested on $\sim$5000 spectra from the Giraffe Inner Bulge Survey (GIBS) that has a spectral window 
similar to that of the Gaia--RVS spectra. Based on this sample, we studied the correlation between the equivalent width (EW) of the 
{\it Gaia DIB} and the interstellar reddening $\EJKs$ toward the inner Galaxy, as well as the DIB intrinsic properties.}
{Our procedure automatically checks and eliminates invalid cases, and then applies a specific local normalization. The
DIB profile is fit with a Gaussian function. Specifically, the DIB feature is extracted from the spectra of late-type stars by subtracting
the corresponding synthetic spectra. For early-type 
stars we applied a specific model based on the Gaussian process that needs no prior knowledge of the stellar parameters. In addition, 
we provide the errors contributed by the synthetic spectra and from the random noise.}
{After validation, we obtained 4194 reasonable fitting results from the GIBS database. An EW versus $\EJKs$ relation is derived as 
$\EJKs\,{=}\,1.875\,({\pm}\,0.152)\,{\times}\,{\rm EW}\,{-}\,0.011\,({\pm}\,0.048)$, according to $\EBV/{\rm EW}\,{=}\,2.721$, 
which is highly consistent with previous 
results toward similar sightlines. After a correction based on the Vista Variables in 
the Via Lactea (VVV) database for both EW and reddening, the coefficient derived 
from individual GIBS fields, $\EJKs/{\rm EW}\,{=}\,1.884\,{\pm}\,0.225$, is also in perfect agreement with literature values. Based on a subsample 
of 1015 stars toward the Galactic center within $-3^{\circ}\,{<}\,b\,{<}\,3^{\circ}$ and $-6^{\circ}\,{<}\,l\,{<}\,3^{\circ}$, we 
determined a rest-frame wavelength of the {\it Gaia DIB} as 8620.55\,{\AA}.}
{The procedures for automatic detection and measurement of the {\it Gaia DIB} are successfully developed and have been applied to
the GIBS spectra. A Gaussian profile is proved to be a proper and stable assumption for the {\it Gaia DIB} as no intrinsic asymmetry is found.
A tight linearity of its correlation with the reddening is derived toward the inner Milky Way, which  is consistent with previous results.}

\keywords{ISM: lines and bands -- 
         dust, extinction --
         Galaxy: bulge
         }
\maketitle
%

\section{Introduction}
Diffuse interstellar bands (DIBs) are a set of absorption features that were first discovered in 1919 \citep{Heger1922}.
These features originate in the interstellar medium \citep[ISM;][]{Merrill1934,Merrill1936} and usually contain broader 
widths than typical atomic lines \citep{Herbig75,Hobbs08}. \citet{Herbig75} was the first to systematically discuss the 
behavior of 39 DIBs in the region of 4400--6850\,{\AA}. An extended search was made by \citet{Sanner78} from 6500 to 8900\,{\AA}.  
\citet{Jenniskens94} made a systematic search for the DIBs on the spectra of four reddened early-type stars and presented 
a catalog containing 229 DIBs, of which 133 were newly detected. The total number of the DIBs increases with the quality 
and wavelength coverage of the spectra. The recently released Apache Point Observatory Catalog contains more than 500 DIBs covering 
optical and near-infrared (NIR) bands \citep{Fan19}. More than 100 years have passed since the first discovery of DIBs, but we still 
know very little about their carriers. The correlation between the strength of the DIBs and interstellar extinction is a 
general property for many strong DIBs \citep{Sanner78,Lan15}. However, the lack of a linear polarization in strong DIBs \citep{Cox07} 
and their missing link to far-ultraviolet extinction \citep{Desert95,XiangFY17} result in the thought that large carbonaceous molecules 
in the gas phase rather than small dust grains are the carriers of the DIBs, for example, polycyclic aromatic hydrocarbons
\citep[PAHs,][]{LH85,ZA85,Salama99,CS06} and fullerenes \citep{Kroto88,Campbell15}. The Buckminsterfullerene, $C^{+}_{60}$, 
is the first and only identified DIB carrier for four DIBs $\lambda 9365$, $\lambda 9428$, $\lambda 9577$, and $\lambda 9632$, 
according to the match of the band wavelengths and the strength ratios between observational and laboratory data 
\citep{Campbell15,Campbell18,Lallement18,Cordiner19}. 

Because DIBs are weak and easily blended with stellar lines \citep{Kos13}, early works preferred high-quality early-type stars 
with only several to a few hundred observations at best. During the past ten years, the upcoming large spectroscopic surveys 
opened a new era in the DIB research, with a considerable number of spectra that allowed constructing a three-dimensional (3D) 
map of the DIBs and unveiling kinematic information and statistical properties of their carriers. Using the spectra from the  
{\it Gaia}--ESO Spectroscopic Survey \citep{Gilmore12}, \citet{ChenHC13} and \citet{Puspitarini15} first detected the DIBs in 
the spectra of late-type stars with automated techniques by fitting the observed spectrum with a combination of a synthetic 
stellar spectrum, a synthetic telluric transmission, and empirical DIB profiles. In addition to the use of synthetic spectra, 
\citet{Kos13} developed a method for detecting interstellar DIBs on cool-star spectra using artificial 
templates constructed from real spectra at high latitudes that are morphologically similar to the target spectrum. 
This method requires no prior knowledge of stellar parameters but can only be applied with large databases. Kos and collaborators 
applied the method to study the DIB around 8620\,{\AA} with $\sim$500,000 spectra from the Radial Velocity Experiment 
\citep[RAVE;][]{Steinmetz06} and built a pseudo-3D map of the DIB strength covering about 3\,kpc from the Sun with a spatial 
resolution between 0.075 and 0.8\,kpc \citep{Kos14}. \citet{YuanHB12} also reported the detection of two optical DIBs $\lambda$5780 
and $\lambda$6283 in about 2000 low-resolution spectra ($R\,{\sim}\,2000$) from the Sloan Digital Sky Survey \citep[SDSS;][]{Eisenstein11}. 
By stacking thousands of SDSS spectra of stars, galaxies, and quasars, \citet{Lan15} successfully created an intensity map of 20 DIBs 
covering $\sim$5000\,$\rm deg^2$ and measured their correlations with various ISM tracers (atomic, molecule, and dust). The tight 
correlation between the strength of the DIBs and interstellar extinction was confirmed toward substantial sightlines. 
The strong DIB at $\lambda\,{=}\,1.527\,\mu m$ (i.e., APOGEE DIB) was thoroughly studied using data from the Apache Point 
Observatory Galactic Evolution Experiment \citep[APOGEE;][]{Majewski16} by \citet{Zasowski15}, \citet{Elyajouri16}, and \citet{Elyajouri19}.
In addition to the common correlation between its strength and extinction, various properties were investigated based on
the large number of APOGEE spectra: \citet{Zasowski15} derived the velocity curve of the DIB carrier and estimated the
rest-frame wavelength of the APOGEE DIB; \citet{Elyajouri19} revealed the depletion of the DIB carrier in dense clouds.

Based on large sky survey projects and new techniques, strong DIBs are identified to be a powerful tool for ISM tomography 
and consequently can probe the Galactic structure, although the carriers are unknown. The forthcoming third data release of the
ESA {\it Gaia} mission that will contain the parameterization of several million spectra will be a leap forward in the sky coverage and spatial
resolution of the DIB intensity map. These spectra are observed with the {\it Gaia} Radial Velocity Spectrometer
\citep[RVS;][]{Recio-Blanco16,Brown2018,Katz2019} for stars as faint as $G\,{\sim}\,15.5$\,mag, with a spectral window from 847 to 871\,nm 
at a resolution of $\sim$11,200. DIB $\lambda$8620 is also the strongest DIB covered by the Gaia--RVS spectra,  known as the ``{\it Gaia DIB}''. 
It was first reported by \citet{Geary75} and has been widely studied for its correlation with interstellar
extinction \citep{Sanner78,Munari00,Wallerstein07,Munari08,Kos13,Damineli16}. Its carrier is not associated with dust grains \citep{Cox11} 
and is still not identified. 

In this paper, we describe our automatic procedure for the detection and measurement of the {\it Gaia DIB,} which can be applied 
for large spectroscopic surveys such as the forthcoming Gaia DR3 release. We applied this method to nearly 5000 spectra from the 
Giraffe Inner Bulge Survey \citep[GIBS;][]{Zoccali14} located in highly extincted regions. The full procedures of the DIB measurements, 
as well as the error analysis, are presented in Sect. \ref{sec:procedure}. The GIBS data are introduced in Sect. \ref{sec:gibs}. 
Section \ref{sec:result} shows the fitting results and the related discussions. Our main conclusions are summarized
in Sect. \ref{sec:conclusion}.


\section{Procedures of the DIB measurement} \label{sec:procedure}

\begin{figure*}
  \centering
  \includegraphics[width=16.8cm]{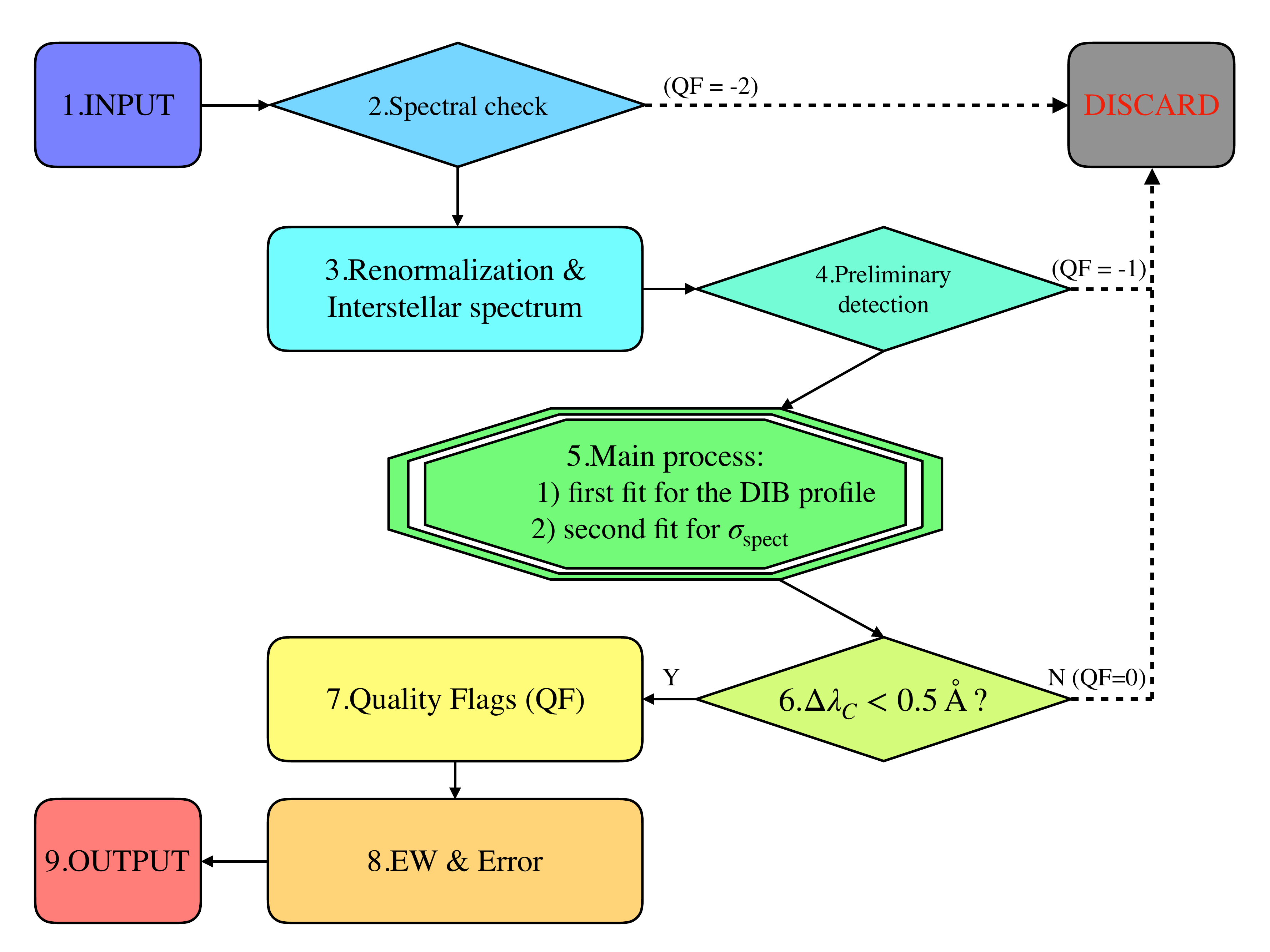}
  \caption{Flowchart compiling our full procedures of the detection and measurement of the {\it Gaia DIB}.}
  \label{fig:flowchart}
\end{figure*}

Most of the DIB studies have focused either on late-type stars (e.g., \citealt{Kos13}) or on early-type
stars (e.g., \citealt{Munari08}) with a reasonable number of spectra to treat (several tens of 
thousands of stars). The challenge of this work is to implement a procedure that is valid for a wide temperature range
and applicable to very large spectral surveys such as that of Gaia RVS. This requires a set of automatic procedures  
that has to be fast in terms of computing time and also reliable. Figure \ref{fig:flowchart} shows the
flowchart of our full procedures, and we describe our automatic procedures in detail below.

\subsection{Inputs and spectral check}

The global inputs of our procedure are the observed spectra corrected for their radial velocities together with their
best-fit synthetic spectrum and the corresponding stellar parameters. We used these stellar parameters together 
with the corresponding synthetic spectra for stars with temperatures from 3500\,K to 7000\,K,
which we call cool stars. We chose this limit in order to ensure that we did not encounter problems 
with the synthetic spectra at the border of their grid. For stars above 7000\,K, which are called hot stars,
we used a specific technique based on the Gaussian process that does not require synthetic spectra, as described in \citet{Kos17}.

The input spectra and parameters (effective temperature $\Teff$, signal-to-noise ratio S/N, radial velocity of the
target star $\rm RV_{star}$ , and its uncertainty $\rm \sigma (RV_{star})$) were checked before further processes to 
eliminate invalid cases. Cool-star 
spectra should have nonzero flux for both observed and synthetic spectra, while for hot-star spectra, only a nonzero observed
flux is required. Stars with $\Teff\,{<}\,3500$\,K were discarded because those spectra are mainly dominated by molecular 
lines that cannot easily be reproduced well by synthetic spectra. In addition, in order to avoid fitting
random-noise profiles instead of the true DIB profiles, we restricted our analysis to stars with $\rm S/N\,{>}\,50$. We  
describe in Sect. \ref{subsubsec:snoise} the effect of the S/N on the error in the DIB measurement. $\rm RV_{star}$ was used 
to convert the central wavelength measured in the stellar rest frame into the heliocentric frame. Targets with large radial 
velocity errors ($\rm \sigma (RV_{star}) > 5\,\kms$) were discarded as well.

\subsection{Interstellar spectra and renormalization}

The interstellar spectra were derived by dividing the observed spectra by the corresponding synthetic spectra 
for cool stars. For hot stars, the {\it Gaia DIB} is usually not blended with stellar lines and can be directly  
measured on the observed spectra, while for cool stars, the stellar lines first need to be removed by using the synthetic spectra.     
We refer here to the cool interstellar spectra and to the hot interstellar spectra as CIS and HIS, respectively.

\begin{figure}
  \centering
  \includegraphics[width=8.6cm]{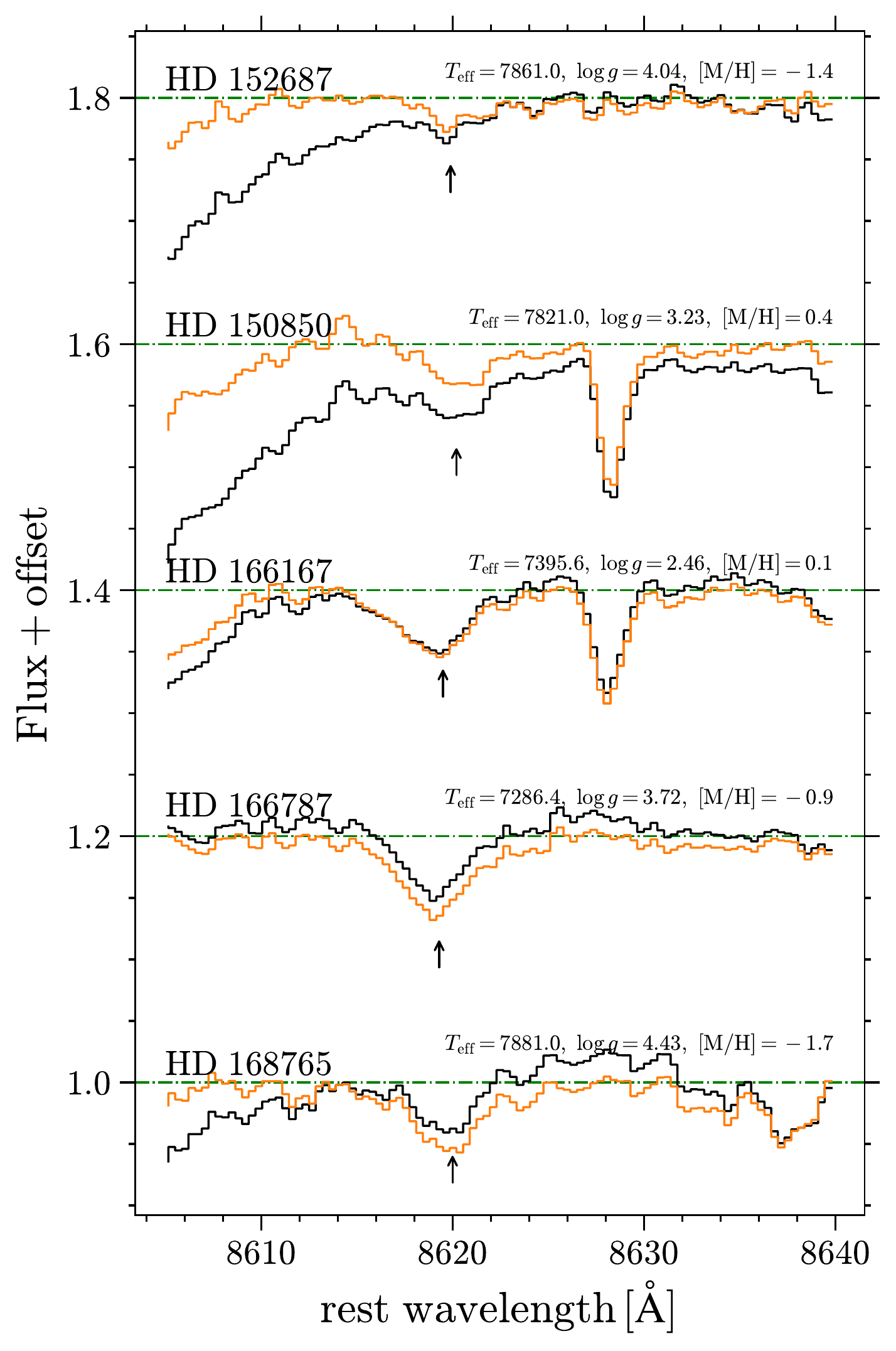}
  \caption{Examples of the local renormalization for five RAVE spectra ($R\,{=}\,7500$) of hot stars.
           The black lines are the original spectra, and the orange lines indicate the renormalized
           spectra. The black arrows indicate the {\it Gaia DIB}. The names and atmospheric parameters are 
           also indicated. The strong stellar feature near the DIB profile seen on HD 150850 and
           HD 166167 is the $\NI$ line.}
  \label{fig:renorm-hot}
\end{figure}

We analyzed and measured the {\it Gaia DIB} in a 35\,{\AA} wide region around its central wavelength 
\citep{Jenniskens94,Galazutdinov00,Munari08}, that is, 8605--8640\,{\AA}. Although the input spectra should be normalized, 
the interstellar spectra usually do not have uniform continua. Especially for hot-star spectra,
heavily uneven continua can be found with the strong hydrogen Paschen 14 line (see Fig. \ref{fig:renorm-hot} 
or the examples shown in \citealt{Munari08}). Therefore, a specific renormalization technique was 
applied to the local spectra within the 8605--8640\,{\AA} spectral window.
For HIS, the local spectrum was first fit by a $\text{second}^{\rm }$-order polynomial, where the differences
of the flux of each pixel to the fitting curve were calculated, as well as their standard deviation. Pixels far away
from the fitting curve were replaced by the corresponding points on the fitting curve. Specifically, for the
pixels above the polynomial, they were replaced when their distances were larger than five times the standard 
deviation. When the pixel was below the fitting curve, the threshold was 0.5 times the standard deviation. 
Different rejected thresholds were set to ensure that the fitted continuum can access the real continuum
and is not lowered by the stellar and/or DIB features. 
The remaining pixels, together with the points replacing outliers, were fit again by a $\text{second}^{\rm }$-order 
polynomial. After 20 iterations, the final fitted polynomial was used as the continuum to renormalize the 
original local spectrum. Figure \ref{fig:renorm-hot} illustrates the local renormalization with five RAVE 
spectra of hot stars. The spectra and their atmospheric parameters were taken from RAVE--DR6 \citep{Steinmetz2020b,Steinmetz2020a}.
The curvatures caused by the Paschen 14 line are alleviated, but the 
DIB and stellar features are kept.

The same technique was also applied to the cool-star spectrum, but using a linear form. The local renormalization and 
the derivation of CIS were made simultaneously following these steps:

\begin{enumerate}
  \item Derive a rough interstellar spectrum, $R_{\rm rough}=F_\lambda / S_\lambda$, where $F_\lambda$ is the observed spectrum and 
        $S_\lambda$ is the synthetic spectrum. $F_\lambda$ and $S_\lambda$ have the same spectral samplings.
  \item Renormalize $R_{\rm rough}$ and extract its continuum, $F_{\rm cont}$.
  \item Renormalize the observed spectrum, $F_{\rm norm} = F_\lambda / F_{\rm cont}$.
  \item Derive the final interstellar spectrum: $R_\lambda = F_{\rm norm} / S_\lambda$.
\end{enumerate}

A renormalized CIS is shown in Fig. \ref{fig:fit-cool} with the corresponding fit of the DIB feature on it.
The spectrum and stellar parameters come from RAVE--DR6 as well.

\subsection{Preliminary detection} \label{subsec:pre-detect}

In order to process a large number of spectra, the fitting of the DIB profile was completely 
automated without any visual inspection. We therefore made a preliminary detection of the DIB profile 
to produce initial guesses for the fitting and eliminated cases whose noise is at the level of or exceeds 
the depth of the DIB feature. The detection was made within the wavelength range between 8614.3--8625.7\,{\AA} 
according to a radial velocity of $\pm 200\,\kms$ of the DIB carrier at the stellar frame. This is a 
reasonable assumption if the DIB carrier mainly traces the local ISM at several kiloparsecs from the Sun.
When the largest depth of the spectrum in this region is larger than $3\,{\times}\,\frac{1}{S/N}$, 
we considered this DIB as a true detection, and the considered spectrum entered the main process of the DIB profile fitting (see 
Fig. \ref{fig:flowchart}), where the depth and its according position were used as the initial conditions 
of $D$ and $\lambda_C$ in the DIB fitting. Otherwise, the case was discarded. 

\subsection{Main process: Fitting the DIB profile} \label{subsec:dib-fit}

The observed profile of the {\it Gaia DIB} along the line of sight could be the superposition of several features  
with different widths but at almost the same wavelength \citep{Jenniskens94}, which can be described by a 
Gaussian profile \citep{Kos13}. We decided to fit the DIB feature on the spectra with a 
Gaussian profile because 1) previous studies revealed no intrinsic asymmetry of the {\it Gaia DIB} \citep{Munari08,Kos13,Puspitarini15}, 
2) the departures from a Gaussian profile caused by the multiple cloud superposition is smaller than other sources of
uncertainty \citep{Elyajouri16}, and 3) a Gaussian fit is easier, more stable, and faster in terms of computing time
than the asymmetric Gaussian fit. 

\subsubsection{Models for the cool and hot Interstellar spectra} \label{subsubsec:fit-models}

The DIB profiles on CIS and HIS were fit by different techniques that are described below.
CIS was modeled using a Gaussian function that describes the DIB profile and a constant
that accounts for the continuum,

\begin{equation} \label{eq:Gauss-fit}
    f_{\Theta}(x;D,\lambda_C,\sigma) = D \times 
    {\rm exp}\left(-\frac{(x-\lambda_C)^2}{2\sigma^2}\right) + C,
\end{equation}

\noindent where $D$ and $\sigma$ are the depth and width of the DIB profile, $\lambda_C$ is the measured central 
wavelength, $C$ is the constant continuum, and $x$ is the spectral wavelength.

However, a simple Gaussian model is not suitable for HIS because they are usually distorted by the strong Paschen 
13 and 14 lines and sometimes contain a strong $\NI$ line around 8629\,{\AA} (see, e.g., HD 150850 and HD 167745 in Fig. 
\ref{fig:renorm-hot}). To fit the DIB profile together with the distorted continuum and possible stellar lines, we 
applied a similar method as in \cite{Kos17} using the Gaussian process (GP) described in detail below. 

The GP is defined as a collection of random variables, any finite number of which have a joint multivariate Gaussian distribution
\citep{Schulz18}. Formally, let the input space be $\mathcal{X}$, and $f$ denotes a function mapping the input space to reals: 
$f: \mathcal{X} \rightarrow \mathbb{R}$. Then, $f$ is a GP if for any vector of inputs $\mathbf{x} = [x_1,x_2,
\ldots,x_n]^{T}$ such that $x_i\,{\in}\,\mathcal{X}$ for all $i$, the outputs $f(\mathbf{x}) = [f(x_1),f(x_2), \ldots,f(x_n)]^T$ is 
Gaussian distributed. GP is specified by a mean function $m(\mathbf{x})$ reflecting the expected function value at input $\mathbf{x}$,
and a kernel (also called covariance function) $k(\mathbf{x},\mathbf{x}^{\prime})$ models the dependence between the output values 
at different input points \citep{Schulz18}. GP can be used as a supervised learning technique for classification and regression.

Gaussian process regression (GPR) is a nonparametric Bayesian approach to regression problems \citep{GB12}. The output $y$ of 
a function $f$ at input $\mathbf{x}$ can be written as 

\begin{equation}
    y = f(\mathbf{x}) + \epsilon
,\end{equation}

\noindent where $\epsilon\,{\sim}\,\mathcal{N}(0,\sigma^2)$ represents the observational error. 
 $f(\mathbf{x})\,{\sim}\,\mathcal{GP}(m(\mathbf{x}), k(\mathbf{x},\mathbf{x}^{\prime}))$ is distributed as a GP \citep{Schulz18}. 
GPR can capture many different relations between inputs and outputs by using a theoretically infinite number of parameters \citep{Williams98}.

For HIS, our goal is to apply GPR to fit the DIB profile and the remaining spectrum simultaneously. The prior mean function is 
often set to $m(\mathbf{x})\,{=}\,0$ in order to avoid expensive posterior computations. Because we wish to extract the information 
of the DIB feature, however, a Gaussian mean function (Eq. \ref{eq:Gauss-fit}) is applied with $C\,{\equiv}\,1$. For the kernels, we followed the 
strategy of \citet{Kos17}: the exponential-squared kernel models the stellar absorption lines,

\begin{equation}
   k_{se}(x,\xp) = a~{\rm exp}\left(-\frac{||x-\xp||^2}{2l^2}\right),
\end{equation}

\noindent and a Mat\'ern 3/2 kernel models the correlated noise,

\begin{equation}
    k_{m3/2}(x,\xp) = a\left(1+\frac{\sqrt{3}||x-\xp||}{l}\right)
    {\rm exp}\left(-\frac{\sqrt{3}||x-\xp||}{l}\right),
\end{equation}

\noindent where $a$ scales the kernels, and $l$ is the characteristic width of each
kernel. 

In principle, the fitting technique based on GP can be applied to spectra of both hot and cool stars. Because it is 
computationally expensive, however, we only applied it to hot-star spectra, which take only a small fraction of the substantial
spectra in large spectroscopic surveys such as Gaia RVS. Nevertheless, as an illustration of this method, we applied it to GIBS
data (see Sect. \ref{sec:gibs}).

\subsubsection{Parameter optimization and MCMC fit}

Maximum likelihood estimation was used to optimize the parameters in the Gaussian model for CIS, that is, $\Theta = \{D, \lambda_C, \sigma, C\}$. 
Given the spectrum $\{\mathbf{X},y,\sigma_y^2\},$ where $\mathbf{X}$ is the wavelength, $y$ is the flux, and $\sigma_y^2$ is the
observational uncertainties (if $\sigma_y^2$ is not accessible, it was fixed to 0.001), the log marginal likelihood is

\begin{equation}\label{eq:ln-like}
{\rm ln}p({y|\mathbf{X}},\Theta) = -\frac{1}{2}\mathbf{r}^T {\textbf{\em K}}^{-1} 
\mathbf{r} - \frac{1}{2}{\rm ln\,det}(\textbf{\em K}) - \frac{N}{2}{\rm ln}(2 \pi),
\end{equation}

\noindent where $\mathbf{r}= {y}-f_{\Theta}(\mathbf{X})$ is the residual vector and $f_{\Theta}$ is the Gaussian model. 
$N$ is the pixel size of the spectrum.
$\textbf{\em K}$ is the covariance matrix,

\begin{equation}
    K = \begin{pmatrix}
        \sigma_1^2 & 0 & \cdots & 0 \\
        0 & \sigma_2^2 & \cdots & 0 \\
        \vdots & \vdots & \ddots & \vdots \\
        0 & 0 & \cdots & \sigma_N^2
        \end{pmatrix}
.\end{equation}

To implement GPR for HIS, we optimized five parameters, three for the DIB profile ($D$, $\lambda_C$, $\sigma$), and two 
for the kernels ($l_{se}$ and $l_{m3/2}$). The scaling factor $a$ of the kernel can be estimated as the variance of the noise and does not need 
to be fit. We used the square of the inverse of the S/N to approximate it. The optimal parameters were estimated by maximizing the 
type \uppercase\expandafter{\romannumeral2} maximum likelihood \citep{RW06}. Its log marginal likelihood is almost the same as Eq. \ref{eq:ln-like}, 
but the covariance matrix becomes nondiagonal,

\begin{equation}
    K_{ij} = \sigma^2_i \delta_{ij} + k(x_i,x_j), ~x \in \mathbf{X},
\end{equation}

\noindent where $\sigma_i$ is the observational error, $\delta_{ij}$ is the Kronecker delta, and $k(x_i,x_j)$ is the element of the specified kernel.

A Markov chain Monte Carlo (MCMC) procedure \citep{Foreman-Mackey13} was performed to implement the parameter estimates for the Gaussian fit and 
GPR. The initial conditions are perturbed by a normal distribution around the initial guess with a standard deviation of 0.01. Different walkers 
of the MCMC can therefore start with different conditions. One hundred walkers were progressed for 50 steps to complete the burn-in stage. The best fits were then used as the 
initial conditions to sample the posterior with 100 walkers and 200 steps. The best estimate and its statistical uncertainty were taken in terms of the
50th, 16th, and 84th percentiles of the posterior distribution.

\begin{figure}
  \centering
  \includegraphics[width=8.6cm]{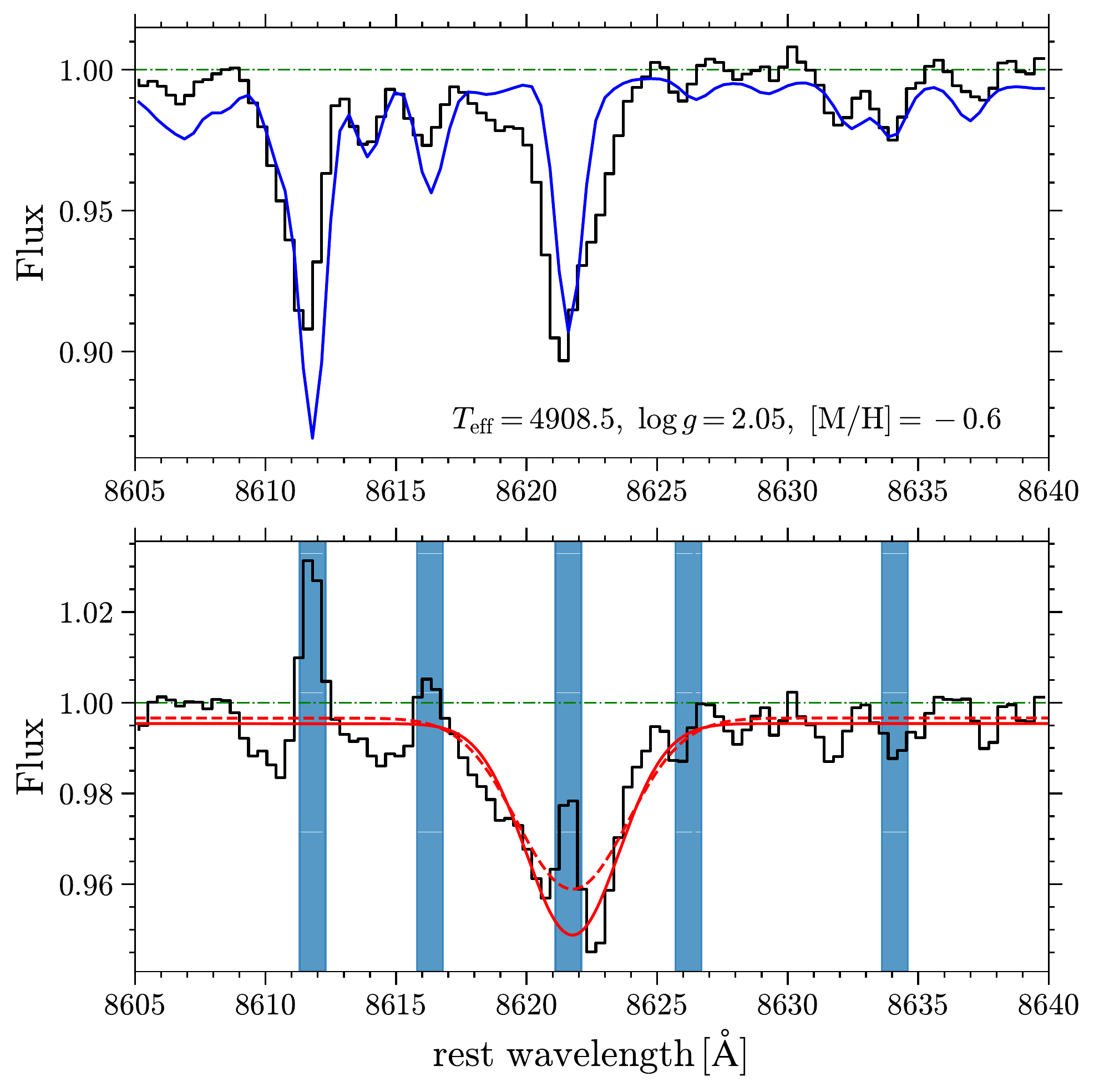}
  \caption{ Gaussian model for CIS of the star HD 149349 observed by the RAVE survey ($R\,{=}\,7500$). Upper panel: Black and 
  blue lines show the observed and synthetic spectra, respectively. The atmospheric parameters from RAVE--DR6 are also indicated. 
  Lower panel: Renormalized interstellar spectrum represented by the black line.
  The dashed and solid red lines represent the profiles from the first and second (finally selected) fits, respectively. 
  The blue shades indicate the masked regions discussed in Sect. \ref{subsubsec:spec-err}.}
  \label{fig:fit-cool}
\end{figure}

\begin{figure}
  \centering
  \includegraphics[width=8.6cm]{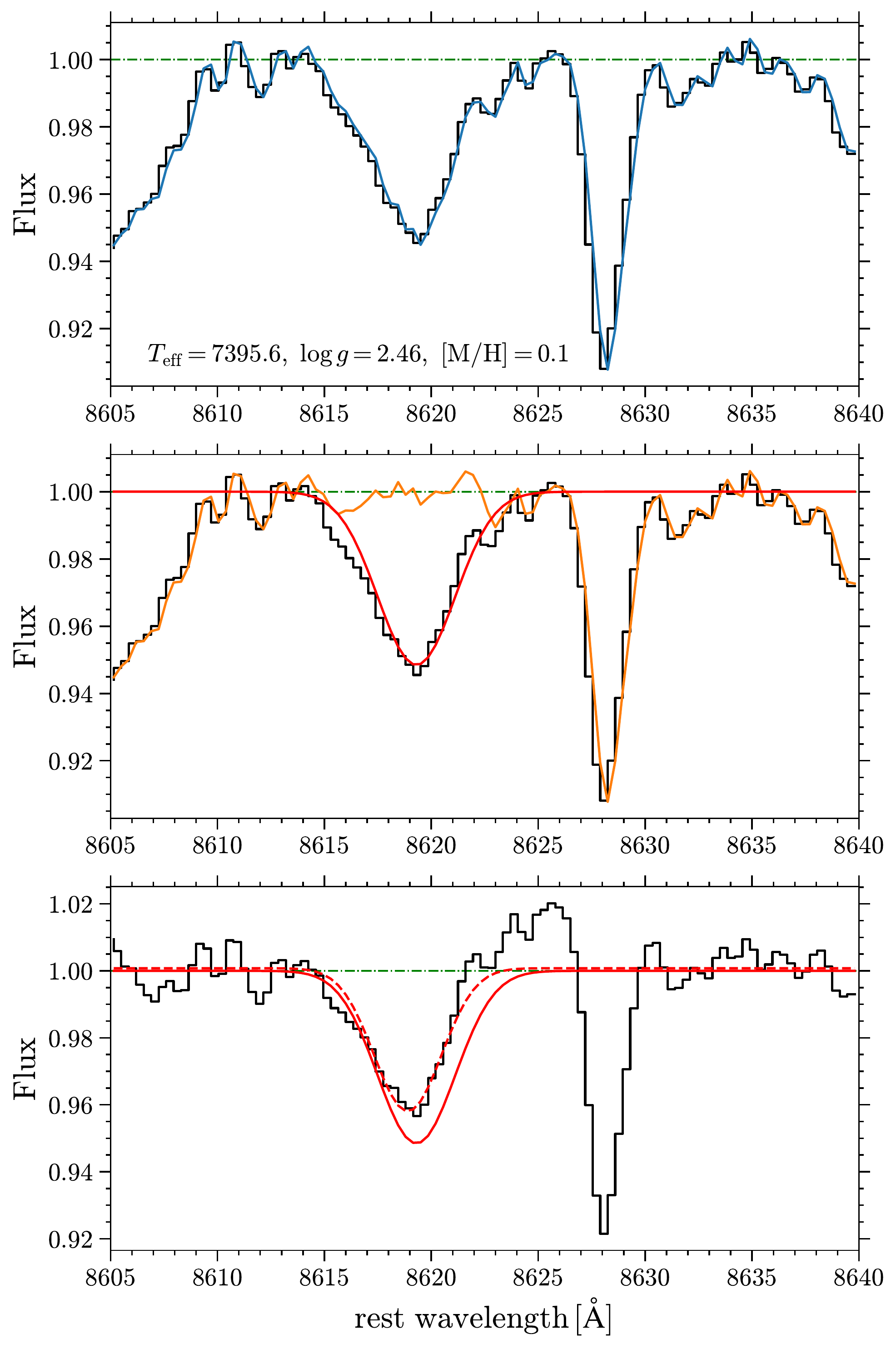}
  \caption{Fitting by GPR for HIS of the star HD 166167 observed by the RAVE survey ($R\,{=}\,7500$). 
  Top: Locally renormalized observed spectrum (black line), and the completed fit by GPR (blue curve). 
  The atmospheric parameters from RAVE--DR6 are also indicated. Middle: Decomposition of the blue curve in the top panel. The red line represents 
  the Gaussian DIB profile, and the orange line given by the kernels of GP describes the remainder of the spectrum. Bottom: 
  Reshaped spectrum based on the first fit (black line; see Sect. \ref{subsubsec:spec-err}). The solid and dashed DIB profiles 
  are from the first (finally selected) and second fits, respectively.}
  \label{fig:fit-hot}
\end{figure}

The initial conditions of $D$ and $\lambda_C$ were measured by the preliminary detection (see Sect. \ref{subsec:pre-detect}). The initial 
values of $l_{se}$ and $l_{m3/2}$ were set to 0.3 and 0.15 for all the cases. $C$ has an initial guess of 1.0 assuming a well-normalized CIS.
The initial guess of $\sigma$ is hard to determine. Based on the GIBS results (see Sect. \ref{sec:result}), $\sigma_0\,{=}\,1.2$ is a proper guess. Strong 
DIB profiles are not sensitive to the initial guess, while weak profiles in general show a good fitting behavior with this value. Examples of the 
DIB fittings for CIS and HIS are shown in Fig. \ref{fig:fit-cool} and \ref{fig:fit-hot}, respectively. We indicate the first initial fit 
and a second fit for the error analysis (see Sect. \ref{subsubsec:spec-err}). The final selected fits are marked as solid lines. 

\subsubsection{Priors} \label{subsubsec:priors}

As a Bayesian approach, priors can be used to prevent unphysical or unreasonable fittings. For the Gaussian fit applied to the CIS, 
we adopted flat priors,

\begin{equation} \label{eq:priors}
    P(D,\lambda_C,\sigma) = \left\{
        \begin{array}{ll}
            1 & if \left\{ \begin{array}{l}
                   0 < |D| < 0.2 \\
                   8610 < \lambda_C < 8627 \\
                   0 < \sigma < 4.0
                   \end{array} \right. \\
            0 & else.
        \end{array}
    \right.
\end{equation}

\begin{figure}
    \centering
    \includegraphics[width=8.4cm]{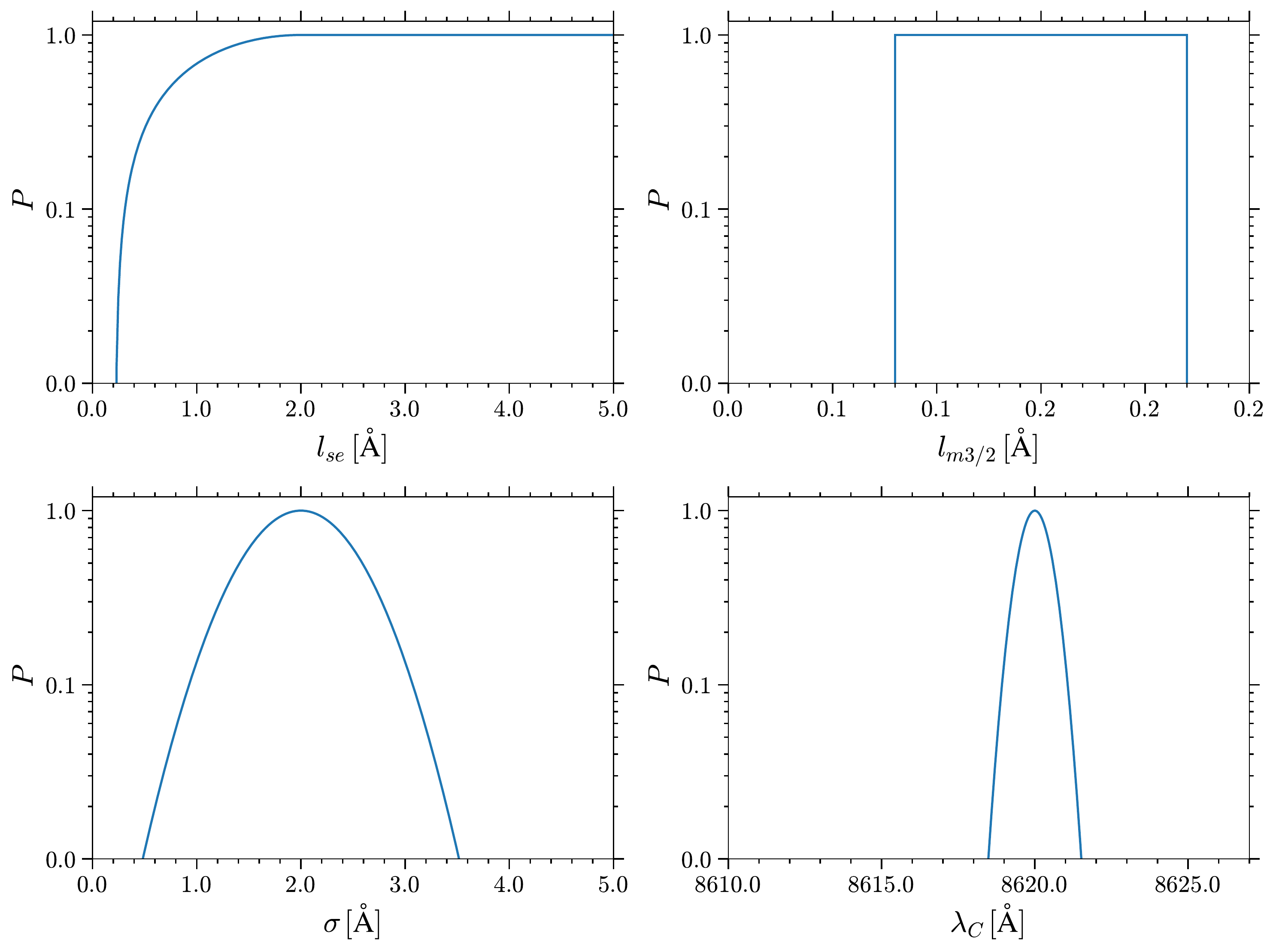}
    \caption{Examples of priors for $l_{se}$, $l_{m3/2}$, $\sigma$, and $\lambda_C$.
             The priors have an original guess of $\sigma_0=2.0$\,{\AA} and $\lambda_C=8620$\,{\AA}.
             This means that $P(l_{se})$ starts at 2.0\,{\AA} to drop to $-\infty$.  $P(\sigma)$ and 
             $P(\lambda_C)$ are centered at 2.0 and 8620 with a width of 0.5\,{\AA}.}
    \label{fig:priors}
\end{figure}

More rigorous priors are needed for HIS to avoid treating the DIB profile as the correlated noise and fit by the kernels of GP. 
The priors of $l_{se}$ and $l_{m3/2}$ provided by \citet{Kos17} were taken. 0.22\,{\AA} was assumed as the boundary of the characteristic 
widths of stellar feature and the random noise, and therefore it is the lower limit of $l_{se}$ and the upper limit of $l_{m3/2}$. $l_{m3/2}$ 
has a lower limit of 0.08\,{\AA}. 
$l_{se}$ is flat at high values and gradually decreases to $-\infty$ at the value of 
the DIB width. $D$ has a flat prior the same as for CIS. The priors of $\lambda_C$ and $\sigma$ are the Gaussian priors centered at their 
initial conditions with a width of 0.5\,{\AA}, which are simpler than those of \citet{Kos17} because they lack the preliminary fit of the DIB 
profile. The prior of $\lambda_C$ is stricter than that of $\sigma$ because its initial guess can be determined by the preliminary 
detection. Some examples of the priors of $l_{se}$, $l_{m3/2}$, $\lambda_C$, and $\sigma$ are presented in Fig. \ref{fig:priors}.

\subsection{Quality flags} \label{subsec:QF}

\begin{figure*}
  \centering
  \includegraphics[width=16.8cm]{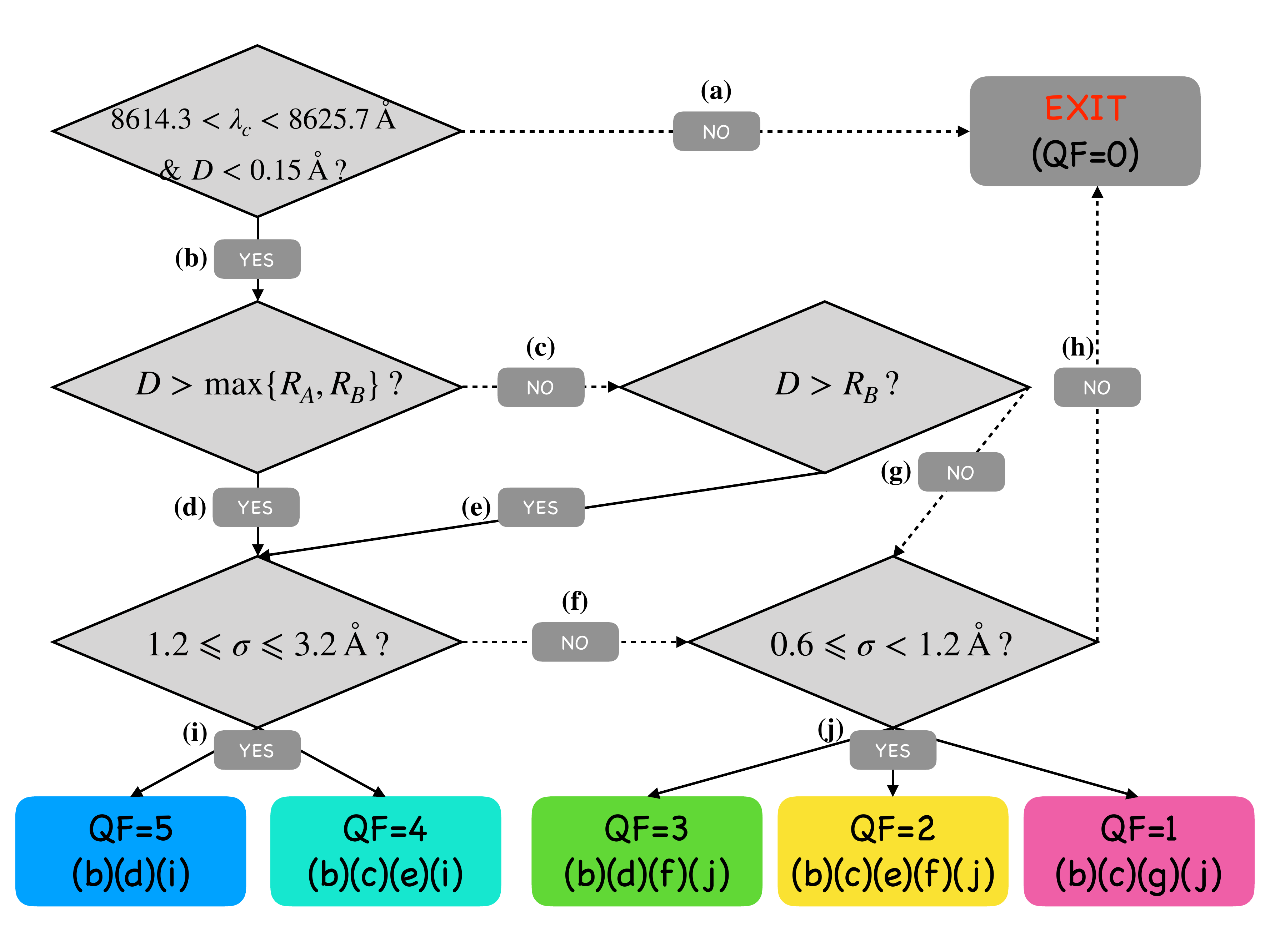}
  \caption{Flowchart of our criteria to generate QF. The flag numbers and
           the corresponding classification paths are listed in the bottom box. Detailed
           explanations are given in Section \ref{subsec:QF}.}
  \label{fig:QF}
\end{figure*}

To select reliable DIB profiles, \citet{Elyajouri16} applied a series of tests to the fit parameters ($D,\lambda_C,\sigma$) and generated 
different quality flags (QF). We follow their main principles. The flowchart of QF is schematically shown 
in Fig. \ref{fig:QF}, and below we describe in detail our procedure to determine QF ranging from $\rm QF\,{=}\,5$ (highest quality) to $\rm QF\,{=}\,0$
(lowest quality). Cases with negative QF were not fit, that is, $\rm QF\,{=}\,{-1}$ was rejected by the preliminary detection, and 
$\rm QF\,{=}\,{-2}$ means invalid spectra.

\begin{enumerate}
    \item {\it Global test:} The first test gives the upper limit of the depth $D$ and 
          the realistic range of the measured central wavelength $\lambda_C$. Here $\lambda_C$ 
          is converted from the stellar frame into the heliocentric frame using $\rm RV_{star}$.
          The cases with $D>0.15$ (the deepest absorption detected on GIBS
          spectra) was eliminated (arrow (a) in Fig. \ref{fig:QF}). These are spurious 
          features generated mainly by the mismatch between the observed and synthetic spectra.
          We also eliminated the fittings with $\lambda_C$ outside the range 8614.3--8625.7\,{\AA,} 
          which is the same interval as we applied in the preliminary detection. 
          
    \item {\it Test of the DIB depth:} When the first test was passed successfully (arrow (b)),
          we compared their depth with the standard deviation of the fitting residuals,
          {\it R\,{=}\,std\,(data--model)}. $R$ was calculated for two regions: $R_A$ is for the
          global spectrum [8605--8640]\,{\AA}, and $R_B$ is in a region close to the DIB
          feature $[\lambda_C-3\sigma,\lambda_C+3\sigma]$\,{\AA}. When $D$ was larger than the
          maximum of $R_A$ and $R_B$ (arrow (d)), then the test on the width was directly
          applied. When the interstellar spectrum is globally too noisy to detect the DIB
          or the DIB is too shallow (arrow (c)), we only compared $D$ with the local
          standard deviation $R_B$. This step allowed us to recover the DIB on the spectra
          that are noisy in some regions far from the DIB, but have good quality near
          the DIB. The cases passing this test (arrow (e)) were subjected to the same test on
          width as the previous ones (arrow (d)). Failed cases (arrow (g)) were 
          examined differently.
          
    \item {\it Test of the DIB width:} In the final test, we defined some limits to select DIBs 
          with reasonable widths. The profiles that exceed the global or local noise level (arrow (d) 
          and (e)) gain high QF with $1.2\,{\leqslant}\,\sigma\,{\leqslant}\,3.2$\,{\AA} or low QF with 
          $0.6\,{\leqslant}\,\sigma\,{\leqslant}\,1.2$\,{\AA}. The shallow DIBs (arrow (g)) were directly tested 
          within the range of 0.6--1.2\,{\AA}. Any case with $\sigma\,{<}\,0.6$\,{\AA} was discarded 
          (arrow (h)) and marked as $\rm QF\,{=}\,0$. Both the lower (0.6\,{\AA}) and upper (3.2\,{\AA}) limits 
          were derived from the GIBS results (see Sect. \ref{subsubsec:dib-para}). 
          Profiles with $\sigma\,{<}\,0.6$\,{\AA} were likely to come from random noise. Extremely 
          broad profiles ($\sigma\,{>}\,3.2$\,{\AA}) are due to unphysical features originating from the data 
          processing or the fitting process. We set 1.2\,{\AA} as the boundary of the two ranges of $\sigma$ 
          to 1) select narrow DIBs through arrow (f) and 2) eliminate the flat and elongated features of 
          uncertain origin \citep{Elyajouri16} for shallow DIBs (arrow (g)).
\end{enumerate}

We created six QFs based on the tests to evaluate the fit quality. $\rm QF\,{=}\,5$ represents the best fits and the detected DIBs with proper 
parameters $\{D,\lambda_C,\sigma\}$. Recovered DIBs with $\rm QF\,{=}\,4$ are locally
detected. Narrow DIBs with $0.6\,{\leqslant}\,\sigma\,{\leqslant}\,1.2$\,{\AA} are flagged 
as 3 or 2 because they exceed the global or local noise level. $\rm QF\,{=}\,1$ corresponds to 
spectra with very low S/N or shallow DIBs. A failed detection is marked as $\rm QF\,{=}\,0$.

\subsection{Equivalent width and error analysis}

The equivalent width (EW) is proportional to the column density of the DIB carriers and reflects the relative oscillator strength 
\citep{Jenniskens94}. With the Gaussian profile, the EW is calculated by the depth $D$ and width $\sigma$,

\begin{equation} \label{eq:ew}
  {\rm EW} = \int \frac{I_0-I_\lambda}{I_0}d\lambda = \sqrt{2\pi}~D~\sigma,
\end{equation}

\noindent where $I_0$ and $I_\lambda$ are fluxes of the continuum and spectrum, respectively. For CIS, the calculated EW was 
further scaled by $C$ because the fit $C$ is usually not unit. There are two main sources of the EW errors, $\sew$, one 
associated with the random noise ($\snoise$), and the other ($\sspect$) contributed by the continuum (HIS) or the mismatches between 
observed and synthetic spectra (CIS). The total error is considered as

\begin{equation} 
  \sew^2 = \snoise^2 + \sspect^2 
.\end{equation}

We estimated $\snoise$ for different DIB profiles by a random-noise simulation.  $\sspect$ was accessed through a second fit for 
each interstellar spectrum. In the following, the estimation of these two errors is explained in detail. 

\subsubsection{Random-noise simulation} \label{subsubsec:snoise}

The random noise discussed in this section mainly refers to the observational uncertainty, which is Gaussian and independent. The random-noise 
level of a spectrum is usually characterized by its S/N. The uncertainty introduced during the data reduction and interstellar spectra 
derivation is discussed and estimated in Sect. \ref{subsubsec:spec-err}. Although the random noise is assumed to be Gaussian, the 
local noise might still distort the DIB profile and affect the fit parameters and consequently the physical quantities such as EW and radial 
velocity. To account for the error of EW contributed by the random noise, \citet{Elyajouri17} made a conservative estimation: 
$\snoise\,{=}\,2\sqrt{2}\,\sigma\,\delta_{\rm depth}$, where $\sigma$ is the fit width of the DIB profile, and $\delta_{\rm depth}$ is the uncertainty of 
the DIB depth. \citet{Puspitarini15} applied a similar formula with a scaling factor $\frac{1}{\sqrt{N}}$, where $N$ is the number of pixels 
covering the DIB width. Their formulas were derived from a series of simulations with varying Gaussian noise and can quickly approximate
$\snoise$. The use of $\sigma$ in their formulas might lead to a strong overestimation for large DIB profiles, however.

For a more comprehensive study and more accurate estimate of the effect of the random noise on the DIB fitting, we performed a series of 
random-noise simulation in the wavelength range between 8605\,{\AA} to 8640\,{\AA} with a pixel size of 0.1\,{\AA}, containing different Gaussian 
DIB profiles ($\{D_0,\mu_0,\sigma_0\}$) and constant continua ($C_0\,{\equiv}\,1$). Then for every spectrum, a Gaussian noise ($\epsilon$) was 
added according to an assigned S/N, that is, $\epsilon\,{\sim}\,\mathcal{N}(0,({\rm S/N})^{-2})$. The parameter grids were constructed as $D_0$ 
ranges from 0.01 to 0.20 with a step of 0.01, $\mu_0\,{\equiv}\,8620.0$\,{\AA}, and $\sigma_0$ ranges from 0.05 to 5.0\,{\AA} with a 
step of 0.05\,{\AA}. For $\rm 20\,{\leqslant}\,S/N\,{\leqslant}\,100$, the step size is 1. For an S/N within 100--300, the step size 5. Higher S/N 
(300--1000) were assigned a step size of 50. Finally, the sample contains 270,000 pseudo-spectra in total with different DIB profiles and S/N. These spectra 
were fit by a Gaussian model with the Levenberg-Marquardt method, and the fit parameters $\{D_f,\mu_f,\sigma_f,C_f\}$ were used to 
study the effect of the random noise. The true ($\rm EW_0$) and fit ($\rm EW_f$) EWs were calculated with Eq. \ref{eq:ew}. 

\begin{figure*}
  \centering
  \includegraphics[width=16.8cm]{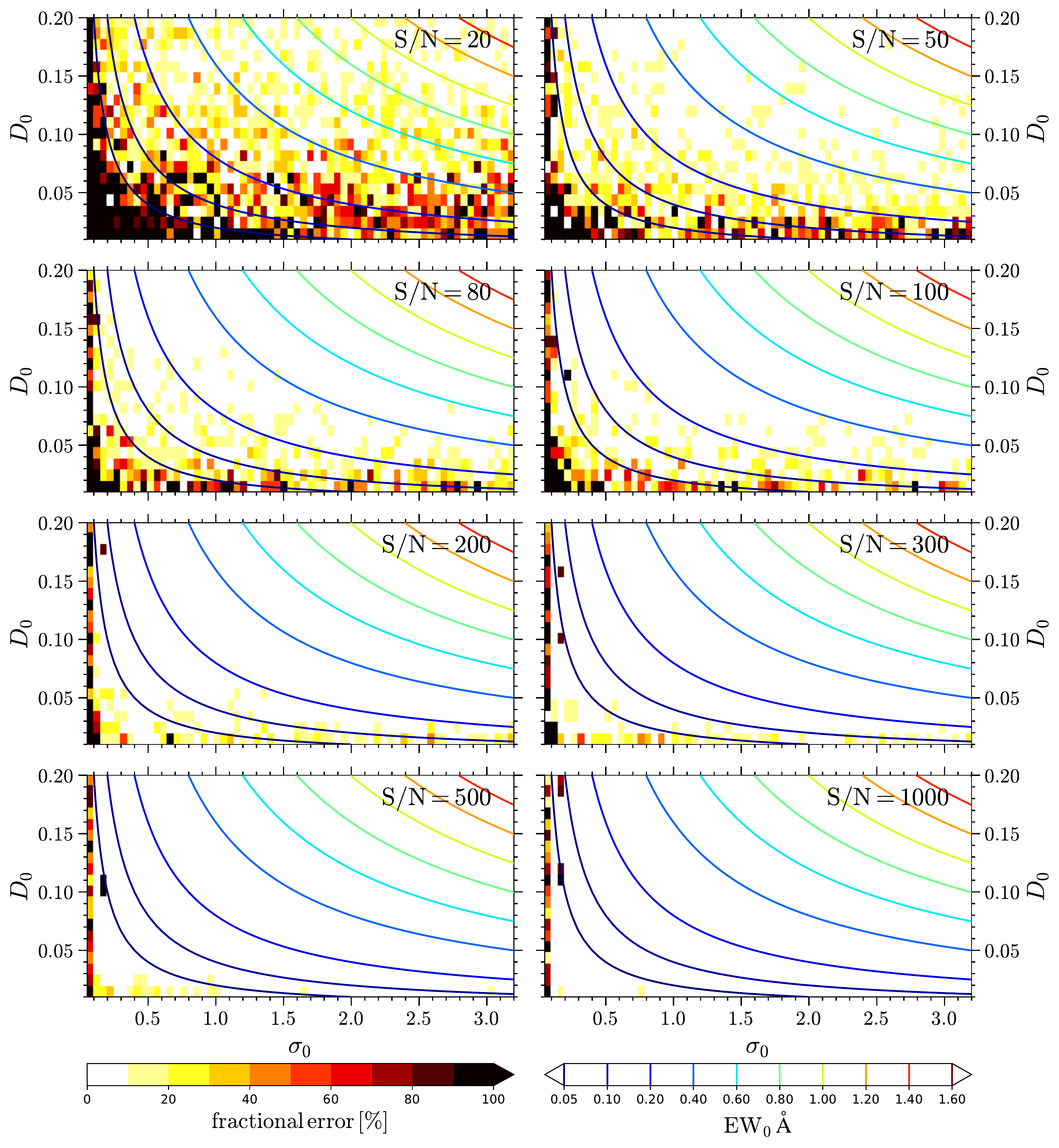}
  \caption{Distribution of the fractional error ($\rm |EW_0-EW_f|/EW_0 \times 100$) in the 
  $D_0-\sigma_0$ plane, overlapped with the contours of $\rm EW_0$. In each subpanel, the fitting
  results are from the pseudo-spectra with the same S/N but various profiles. The color bar is shown 
  within 100\%, while the fractional error may exceed it in some very small $\rm EW_0$ regions.}
  \label{fig:simu1}
\end{figure*}

Figure \ref{fig:simu1} shows the distribution of the fractional error between $\rm EW_0$ and $\rm EW_f$ ($\rm |EW_0-EW_f|/EW_0$) in the $D_0-\sigma_0$ plane for some 
specific S/N, overlapped with some contours of $\rm EW_0$ calculated by the according $D_0$ and $\sigma_0$. The change in fractional errors
is rough because we only fit each spectrum once. The shown $\sigma_0$ is limited to 3.2, the same as the largest valid width detected on GIBS 
spectra. Generally, the fractional error decreases with the increase in S/N and $\rm EW_0$. For $\rm S/N\,{>}\,200$, the fractional errors are smaller than 
10\% (white regions in subpanels in Fig. \ref{fig:simu1}) for most of the fitting results, but for $\rm S/N\,{=}\,100$, $\rm EW_0$ has to be 
as large as 0.4\,{\AA} to ensure that most of the fractional errors are within 10\%. Nevertheless, some shallow profiles could still gain large errors
up to 20\%. If the spectra have $\rm S/N\,{\approx}\,50$, the random noise can cause fractional errors as large as 20\% even for $\rm EW_0\,{>}\,0.5$\,{\AA}, 
which is stronger than most of the {\it Gaia DIBs} detected in previous works \citep{Sanner78,Munari08,Puspitarini15}.
{\it Therefore we only regard sources with S/N higher than 50.} Moreover, for a 
given $\rm EW_0$, shallow DIB profiles tend to gain larger errors than narrow ones because the shallow profiles cover more pixels. If $\sigma_0$ 
is approximate to the pixel size of the spectra, the fractional error maintains a high level and does not significantly decrease with S/N. This also
occurs when $D_0$ is approximate to $\rm \frac{1}{S/N}$. This implies the detection limits for the width and depth of DIB. The effects of the random 
noise on $D_0$ and $\sigma_0$ are similar to that on EW, while the error of $\mu_0$ is more sensitive to $D_0$ than $\sigma_0$. 
The random noise has almost no effect on the continuum $C_0$ for well-normalized spectra. 

Based on the random-noise simulation, the effect of random noise on the DIB fitting was studied in detail. Estimating of $\snoise$ is not 
straightforward, however, because in practice we can only access the fit parameters and not their true values. That is to say, we have to use
$\{D,\mu,\sigma\}$ instead of $\{D_0,\mu_0,\sigma_0\}$ to estimate the error of $\rm EW_f$ contributed by the random noise, that is, $\snoise$.
We try to build a model based on the random forest regression, which is an ensemble machine-learning method combining a large number of decision trees 
\citep{Breiman01}. The model returns $\snoise$ when given $\{D,\sigma,{\rm S/N}\}$. $\mu$ and $C$ were not used because in our simulation $\mu_0$ and 
$C_0$ were fixed. A quarter of the simulation results that were uniformly selected with $\rm EW_0$ constituted the training set, and the test set 
consisted of the remaining part. The regression was completed by the Python scikit-learn package \citep{scikit-learn}. We used 100 trees in the forest 
(n\_estimators=100) and followed the default values of other main parameters. The differences between the true and estimated $\snoise$ are mainly within 
0.05\,{\AA} for the training set and 0.1\,{\AA} for the test set, and they do not significantly change with $\rm EW_0$. The uncertainty for $\rm S/N\,{<}\,50$ 
could be up to 0.2\,{\AA} and higher. The performance of the model is limited by the fact that the features we used $\{D,\sigma,{\rm S/N}\}$ in the 
algorithm are not enough to fully access the true $\snoise$. The estimate of $\snoise$ is accurate for large DIB or high-quality spectra, but 
it is less reliable for small DIB or low-S/N spectra.

\subsubsection{Spectral contribution} \label{subsubsec:spec-err}

To obtain $\sspect$, each spectrum was fit twice. The first fit is detailed in Section \ref{subsec:dib-fit}. 
The second fit considered the effect of the observed--synthetic mismatch for CIS and continuum for HIS. The difference of EWs between 
these two fits was used to estimate $\sspect$.

For CIS, the second fit is still a Gaussian fit, but with five masked regions centered at 8611.8\,{\AA}, 8616.3\,{\AA}, 
8621.6\,{\AA}, 8626.2\,{\AA}, and 8634.1\,{\AA}, with a width of 1\,{\AA} for each of them. These regions correspond to some strong
stellar lines, for instance, the $\FeI$ lines at 8610.602\,{\AA}, 8611.804\,{\AA}, 8613.935\,{\AA}, 8616.276\,{\AA}, 8621.601\,{\AA}, and 8632.412\,{\AA};
the $\CaI$ line at 8633.933\,{\AA}; and the $\TiI$ line at 8618.425\,{\AA} (the strength of these stellar lines would vary with the stellar types 
and metallicities), that may be poorly modeled by the synthetic spectra. 
Figure \ref{fig:fit-cool} shows that the mismatches in the masked regions are higher than average. Consequently, a large $\sspect$ 
is obtained. Although this method is incomplete because we cannot mask all of the abundant stellar lines in the DIB analysis interval, 
it is still a good estimate of $\sspect$ for strong DIBs, as discussed by \citet{Puspitarini15}.

Although the local renormalization corrects for the curved continuum of HIS, some curvatures could still remain and lead to an underestimated EW. 
After extracting the fitted DIB profile, we therefore applied polynomials from $\text{first}^{\rm }$ to $\text{sixth}^{\rm }$ order to fit the remaining 
HIS to approximate the possible curved continuum. Then the best fit was used to renormalize HIS again, and we refit the DIB profile with the 
Gaussian model (Eq. \ref{eq:Gauss-fit}). 

To ensure that the second fit is reasonable, the new DIB profile was preferred only if it was stronger and deeper than the first. Otherwise, 
we retained the results from the first fit. For example, the second fit was accepted for star HD 149349 (Fig. \ref{fig:fit-cool}), 
while it was rejected for star HD 166167 (Fig. \ref{fig:fit-hot}). Furthermore, we obtained an EW of $0.234\,{\pm}\,0.039$\,{\AA} for star HD 
166167, which is consistent with the value of 0.217\,{\AA} reported by \citet{Munari08}, who applied a$\text{}^{\rm }$ sixth-order polynomial to fit the 
continuum and then calculated the EW by integration. Additionally, the cases with large differences between the measured central wavelength 
of the two fits were eliminated, that is, $\Delta \lambda_C\,{>}\,0.5$\,{\AA} (the $\text{sixth}^{\rm }$ step in Fig. \ref{fig:flowchart}). These cases 
were also marked $\rm QF\,{=}\,0$.

\subsection{Outputs and summary}

The final output of each fitting was included in the fit parameters (
$\{D,\lambda_C,\sigma,C\}$ for CIS, $\{D,\lambda_C,\sigma,l_{se},l_{m3/2}\}$ for HIS),
calculated EW and its errors ($\sew$, $\snoise$, $\sspect$), and QF.
For discarded cases, all the parameters were set as {--1}.

\section{Application to the GIBS dataset} \label{sec:gibs}

\subsection{GIBS spectra} 

\begin{figure}
  \centering
  \includegraphics[width=8.4cm]{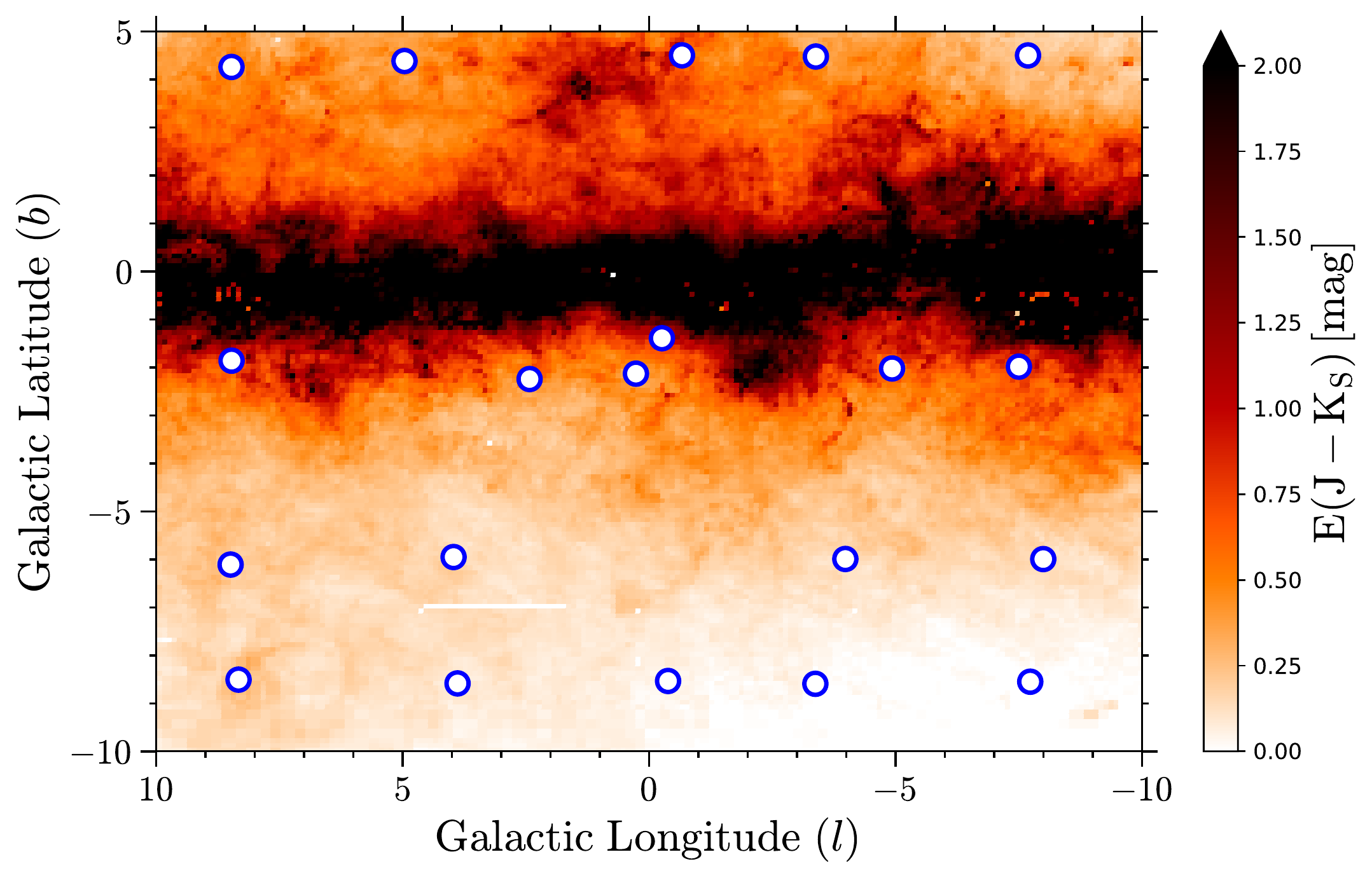}
  \caption{Locations of 20 fields (blue and white circles) in the GIBS survey, overplotted on an extinction 
  map derived by \citet{Gonzalez12}. The locations of all the 31 fields of the GIBS survey can be found in Fig. 1 of
  \citet{Zoccali14}.}
  \label{fig:gibs-survey}
\end{figure}

The GIBS is a survey of red clump (RC) stars selected from the Vista Variables in 
the Via Lactea (VVV) catalogs \citep{Minniti10} in the Milky Way bulge \citep{Zoccali14}. We used 4797
low-resolution spectra from 20 observational fields in the GIBS survey (see Fig. \ref{fig:gibs-survey}). 
The GIBS spectra analyzed here are from the GIRAFFE LR8 setup at the resolution $R\,{=}\,6500$ with the spectral coverage 
of 8206\,{\AA}\,${<}\,\lambda\,{<}$\,9400\,{\AA}. For the analysis, we selected a smaller range of 8450--8950\,{\AA}
because beyond the shorter interval adopted for the analysis, many skylines affect the spectra and we are interested 
in the region around the calcium triplet lines where the {\it Gaia DIB}  is located. We calculated the S/N
of each spectra between 8850\,{\AA} and 8858.5\,{\AA,} where no strong stellar lines are present.  
We also used the templates of RCs to subtract stellar components from the DIB measurement. 
The synthetic spectra used in this work were generated by the Turbospectrum code \citep{AP98}, the MARCS atmosphere model \citep{Gustafsson08}, 
and the line list of the Gaia--ESO survey \citep{Heiter20pre} with $\Teff\,{=}\,4500$\,K, $\logg\,{=}\,2.5$ and the particular metallicity of each star.  
As the residuals in the interstellar spectra are too high because of the mismatch between observed and synthetic spectra,
we did not apply the CIS method to the GIBS data set.

\subsection{Extinction}

Based on the VVV survey, \citet{Gonzalez11,Gonzalez12} built the first complete bulge extinction map (G12 henceforth)
with a differential method. They first derived the mean $(J-K_{\rm S})$ color of the RC stars in 
1835 subfields and then compared it to the color of RCs in a referred region with known extinction, that is, 
Baade's Window. The extinction calculated with the BEAM calculator\footnote{http://mill.astro.puc.cl/BEAM/calculator.php} 
is shown in Fig. \ref{fig:gibs-survey}. The resolution varies from 2{\arcmin} to 6{\arcmin}. With a newly developed 
$JK_{\rm S}$ photometry catalog of the VVV survey based on point spread funciton (PSF) fitting \citep{Surot19}, \citet{Surot20} 
calculated the mean $(J-K_{\rm S})$ of RC+RGB (red giant branch) stars in finer bins. A calibration was then made by comparison with G12 
in areas with $|b|\,{>}\,3^{\circ}$ to derive the absolute extinction values. The improved extinction map (S20 henceforth) 
has a higher resolution that can reach subarcmin  in low-latitude regions.

We derived the extinctions of the GIBS targets from S20 according to their spatial positions. Neither S20 nor 
G12 were able to resolve the extinction of individual GIBS targets for $|b|\,{>}\,3^{\circ}$ because the spatial 
resolution of these maps decrease with increasing latitudes.

\subsection{Specific model for GIBS spectra}

We used the GIBS spectra to validate our procedures of the DIB detection and measurement presented in
Sect. \ref{sec:procedure} and also test the fitting technique based on GPR. We applied the GPR method to fit the 
DIB profile on the derived interstellar spectra from GIBS because 1) the data size is small, therefore the computational
time is acceptable, and 2) the templates with constant $\Teff$ and $\logg$  cause considerable mismatch and
correlated noise that prevent fitting the simple Gaussian model. The simple Gaussian model has been widely
used to fit various DIB profiles on a substantial number of spectra, for example, \citet{Kos13},
\citet{Lan15}, \citet{Zasowski15}, and \citet{Elyajouri17}. 
We therefore did  not select a specific sample to test the Gaussian fit used for cool-star spectra. An illustration
is shown in Fig. \ref{fig:fit-cool}.

Furthermore, we applied both Gaussian and asymmetric Gaussian models to the GIBS spectra to study the amplitude and 
the effect of the asymmetry caused by the velocity dispersion of DIB carriers along different sightlines. We 
chose a simple method presented in \citet{Kos17} to implement the asymmetry, which was introduced by making the
width a function of the wavelength,

\begin{equation}
    \sigma(\lambda; \lambda_C, asym) = \frac{2\sigma}{1 + {\rm exp}(asym \cdot (\lambda-\lambda_C))}
,\end{equation}

\noindent where $asym$ is the amplitude of the asymmetry, and it is limited within $\pm 0.75$
in the fittings. 

\section{Results and discussions} \label{sec:result}

In this section, we study the correlation between EW and reddening, as well as the properties
of the {\it Gaia DIB} based on the fitting results. In general, the Gaussian fit and the asymmetric Gaussian
fit yielded similar profiles (see Section \ref{subsubsec:dib-asym}). We therefore base our results and
discussions on the Gaussian fit alone.

Because of the applied synthetic spectra, the local 
renormalization for the GIBS spectra could lead to nonunit continua and cause an overestimation of 
the EW. We therefore performed a calibration to the EW according to the residual spectra ({\it data--model}).
The fit profiles were shifted according to the mean flux of the residual spectra where the EW were
recalculated as well. Most of the mean flux is within 0.95--1.0, implying an overestimation of the 
EW. For 378 cases the local continua  are above 1.0. After visual inspection, they were eliminated 
because the fit profiles were not physical or too noisy. The cases that did not pass the 
preliminary detection were also discarded. Finally, we obtained 4194 valid GIBS spectra for our DIB analysis. As the average
S/N of GIBS spectra is about 80 per pixel, the fit parameters (${D,\lambda_C,\sigma}$) and EW are 
noise dominated, especially for small DIBs. 

\subsection{Linear correlation between EW and reddening} \label{subsec:ew-ext}

Several studies have revealed a linear correlation between EW and reddening in the optical bands for the {\it Gaia DIB}, 
usually taking the form of $\EBV\,{=}\,a\,{\times}\,{\rm EW}$. Some early estimates 
of $a$ are based on several dozen hot stars, for example, 2.85 \citep{Sanner78} (the coefficient
was calculated by \citet{Kos13}), 2.69 \citep{Munari00}, 4.61 \citep{Wallerstein07}, and 2.72 \citep{Munari08}. The measurement
of $\EBV$ in \citet{Wallerstein07} is doubtful, as discussed in \citet{Munari08}. By merging 
several thousand RAVE cool-star spectra, \citet{Kos13} derived $a\,{=}\,2.49$ with an offset of 0.028. 
\citet{Kos13} also studied 114 RAVE hot stars (including 31 objects from \citet{Munari08}), 
which yielded a highly consistent value of 2.48. \citet{Puspitarini15} also 
derived a linear relation between EW of the {\it Gaia DIB} and $A_0$ (extinction at $\lambda\,{=}\,5500$\,{\AA}) 
toward the Galactic anticenter based on 64 cool stars from $\Gaia$--ESO, but no coefficient
was given. From Fig. 7 in \citet{Puspitarini15}, we estimate a coefficient of $\rm A_0/EW\,8620\,{=}\,2.3/0.35\,{=}\,6.57$. 
Applying the CCM89 model \citep{CCM89}, we have $\rm \EBV/EW\,{=}\,2.12$ ($\Rv\,{=}\,3.1$), a
value significantly lower than others. While \citet{Damineli16} reported a quadratic relation
between $\AKs$ and EW based on $\sim$100 hot stars in and around the stellar cluster Westerlund 1.
Their relation is close to \citet{Munari08} for $\rm EW\,{<}\,0.5$\,{\AA}.
As discussed below, we derived a linear correlation between EW and $\EJKs$ for our GIBS sample by averaging over different 
reddening bins (Sect. \ref{subsubsec:mean-relation}) or individual fields (Sect. \ref{subsubsec:ini-field}).

\subsubsection{EW versus $\EJKs$ relation} \label{subsubsec:mean-relation}

\begin{figure}
  \centering
  \includegraphics[width=8.4cm]{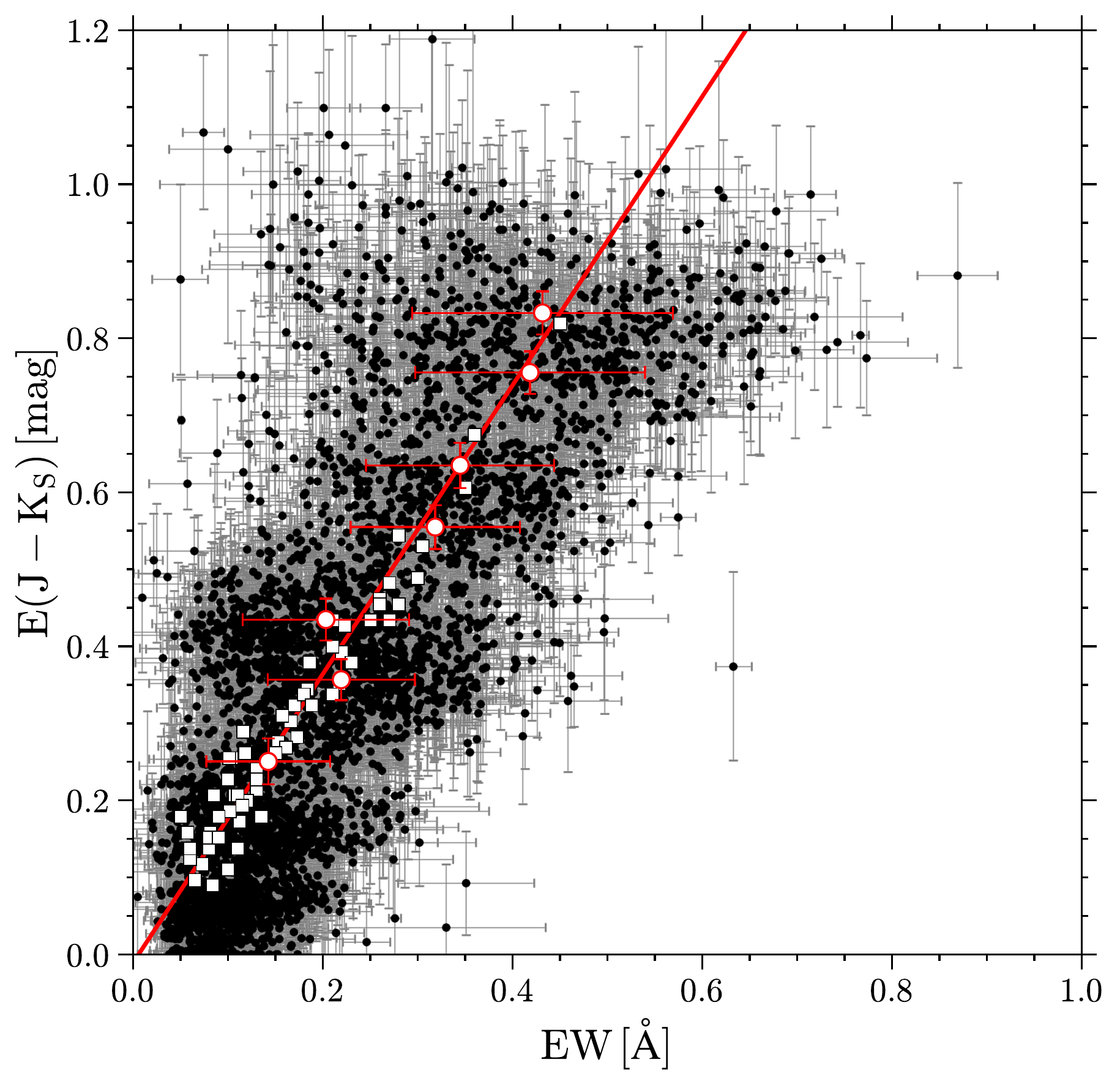}
  \caption{Correlation between the EW of the {\it Gaia DIB} and the reddening $\EJKs$. The black points are the 
  measurements from individual targets. The white and red dots are the median values taken from different 
  reddening bins. The red error bars present the standard deviations in each bins. The red line is fit to 
  the white and red dots. The white squares are the results from \citet{Munari08}.}
  \label{fig:ew-mean}
\end{figure}

The correlation between the EW of the {\it Gaia DIB} and the reddening $\EJKs$ from S20 for individual GIBS targets is shown 
in Fig. \ref{fig:ew-mean}, overlapped with the measurements from \citet{Munari08}. We derived the linear relation over 
the GIBS targets by taking the median values from different reddening bins, ranging from $\EJKs\,{=}\,0.2$ to 0.9 with a 
bin size of 0.1\,mag (the white and red dots in Fig. \ref{fig:ew-mean}). The linear fitting for these median points 
yields $\EJKs\,{=}\,1.875\,({\pm}\,0.152)\,{\times}\,{\rm EW}\,{-}\,0.011\,({\pm}\,0.048)$. The coefficients and their
standard errors were derived with the Python package {\it statsmodels} \citep{python-statsmodels}.

The median points in bins with $\EJKs\,{<}\,0.2$\,mag were not used because they deviate significantly from the linear relation. This 
is because a) spectra containing very small DIBs ($\rm EW\,{<}\,0.05$\,{\AA} before calibration) did not pass the 
preliminary detection because of their low S/N. Therefore the median value of EW is higher than expected. 
b) Most of the small DIBs are noise dominated and might be overestimated by the local increase of noise. 
It is not possible to select and discard the overestimated cases even by visual inspection. Reliable fittings 
with $\rm EW\,{\approx}\,$0.1--0.2\,{\AA} can also be found in low-extinction regions, however.

\begin{figure}
  \centering
  \includegraphics[width=8.4cm]{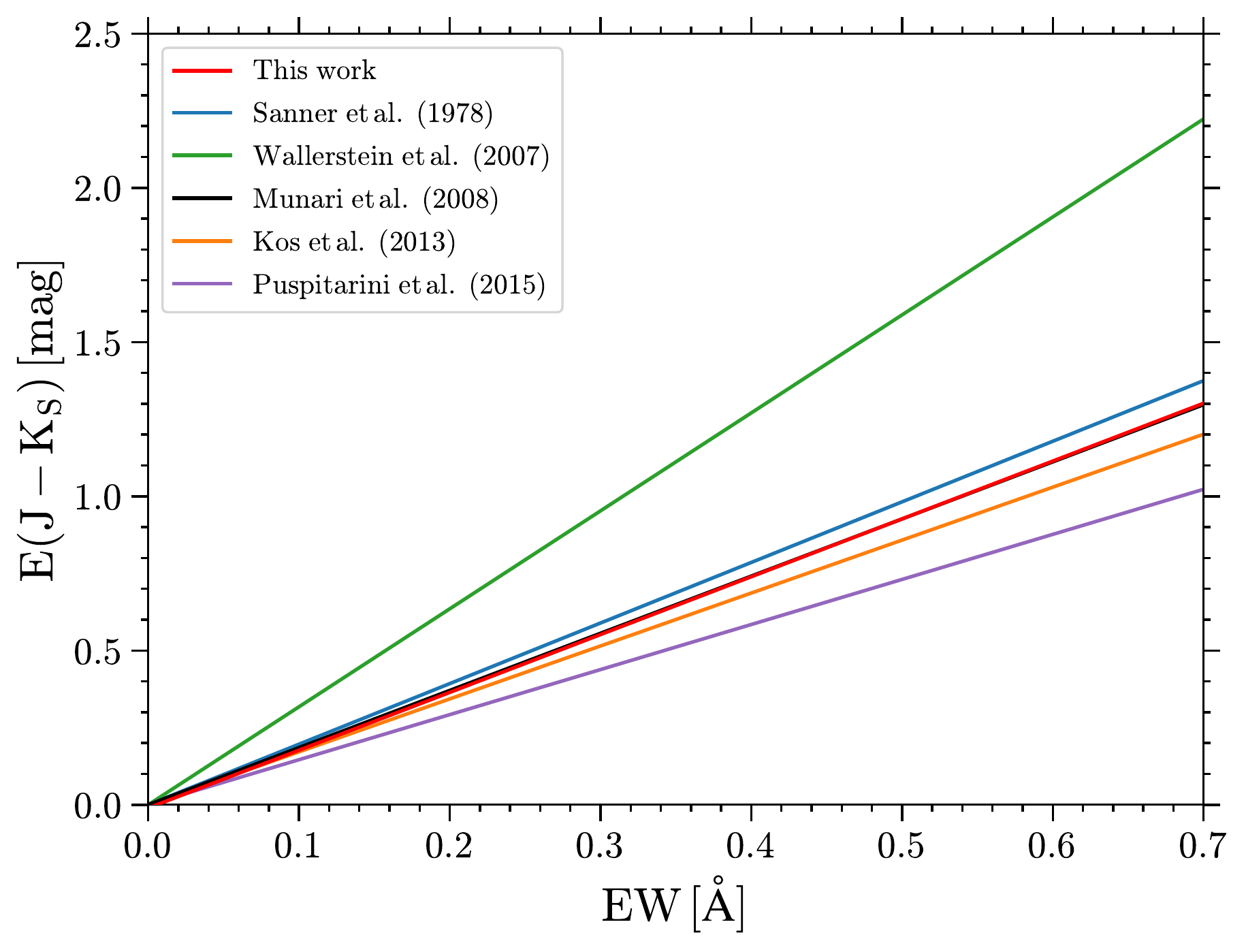}
  \caption{Comparison between the EW--$\EJKs$ linear relations derived from different works. Relations 
  derived in optical bands are translated into the extinction laws applied in Sect. \ref{subsubsec:mean-relation}. 
  The lines of this work (red) and \citet{Munari08} (black) overlap.}
  \label{fig:ew-compare}
\end{figure}

In the absence of optical photometry and in order to compare our results with previous works, a conversion between $\EJKs$ and 
$\EBV$ is needed. For the NIR bands, we applied the extinction law derived by \citet{Nishiyama09} toward the Galactic center: 
$\frac{\AKs}{\EJKs}\,{=}\,0.528$, which is widely used for Galactic Bulge studies, while for the optical--NIR conversion, we still used the CCM89 
model with $\Rv\,{=}\,3.1$ and the corresponding ratio $\frac{\AKs}{\EBV}\,{=}\,0.364$ because 1) \citet{Gonzalez11,Gonzalez12} used the same ratio to
calculate $\EJKs$ for RC stars, which has been used by \citet{Surot20} to calibrate their extinction map. Therefore $\EJKs$ used in this work already
implies a specific ratio between $\AKs$ and $\EBV$. 2) Although many works have pointed out that the extinction law toward
the inner Milky Way deviates from the CCM89 model with $\Rv\,{=}\,3.1$ \citep[e.g.,][]{Indebetouw05,Nishiyama06,Nishiyama09,Nataf16,Damineli16},
the optical--NIR relation is not studied as well as for NIR bands, and the ratio of $\frac{\AKs}{\EBV}$ toward the Galactic
Bulge is still only poorly determined (\citet{Nataf16} did not cover $\EBV$ and \citet{Damineli16} mainly studied
Westerlund 1). With the translation ($\frac{\EJKs}{\EBV}\,{=}\,0.689$), we obtain 
$\frac{\EBV}{\rm EW}\,{=}\,2.721$, which is highly consistent with the result of \citet{Munari08} (see the red line and
white squares in Fig. \ref{fig:ew-mean}). The comparison with other works is shown in Fig. \ref{fig:ew-compare}, and the 
coefficients for the linear relation in the NIR bands are listed in Table \ref{tab:ew}.

The choice of extinction law significantly affects the comparison. Assuming the CCM89 model in NIR bands ($\Rv\,{=}\,3.1$),
$\frac{\AKs}{\EJKs}\,{=}\,0.688$, we have $\frac{\EBV}{\rm EW}\,{=}\,3.544$. With the extinction laws derived by \citet{Damineli16}
from optics to NIR, $\frac{\AKs}{\EBV}\,{=}\,0.297$ and $\frac{\AKs}{\EJKs}\,{=}\,0.449$, we obtain a ratio of $\frac{\EBV}{\rm EW}\,{=}\,2.833$.
Although without the optical photometry for GIBS targets, the  high consistency between optical and NIR relations against
the {\it Gaia DIB} still needs to be confirmed by more studies, our comparison and the consistency found in this work and 
\citet{Damineli16} imply that the correlations of EW with optical and NIR extinctions are at least not very far away from each other. 
However, the relations between EW\,8620 and extinction in different bands and the extinction law have been studied very little.
On the other hand, the correlation between EW and extinction is also related to the dust properties along the line of sight. \citet{Ramrez-Tannus18} 
reported a linear correlation between extinction-normalized EW, $\rm EW/\Av$, and $\Rv^{-1}$ for 14 DIBs in M17. They derived a 
relation between EW and $\EBV$. \citet{Li-Kaijun19} used $\EBV$ to normalize the EW and reported no
relation with $\Rv^{-1}$. They suggested that the hydrogen column density, $N_{\rm H}$, is a more 
appropriate normalization than extinction and reddening. Theoretically, small dust grains would present
a steep extinction curve with small $\Rv$, and very large grains experience a flat curve with $\Rv\,{\to}\,\infty$
\citep{Draine03}. This means that different $\Rv$ values should indicate different correlations between EW and extinction,
especially in the ultraviolet and optical bands. Although the variation in $\Rv$ in a single region is always
smaller than the uncertainty in EW and extinction, we could investigate it from different sightlines: the linear
coefficient between EW and $\EJKs$ toward the Galactic Bulge \citep[this work and][]{Munari00,Munari08} is 
apparently larger than the value toward the Galactic anticenter \citep{Puspitarini15}, although this
difference might also be caused by the dependence of the extinction laws on different lines of sight. \citet{Kos13}
derived a mediate value from a substantial number of sightlines. Nevertheless, our studies are also affected
by the method of extinction calculation and DIB measurement, which undermines the credibility of the variation
of EW--extinction correlation for different sightlines. On the other hand, the environmental dependence of
the DIB carriers complicates this question as well. For example, the deviation from linear relation in high-extinction 
regions ($\Av\,{\approx}\,$10\,mag) may be caused by the carrier depletion in dense cores \citep{Elyajouri19}.
The forthcoming Gaia--RVS spectra are expected to bring new insights into the DIB properties. Its large spatial 
coverage and uniform DIB and extinction measurements will give us an opportunity to unveil the relation
between the DIB strength and the corresponding dust properties. 

\begin{table}
\begin{center}
\caption{Coefficients and their uncertainties of the linear relations between the {\it Gaia DIB} EW and reddenings given in the
literature and this work. Relations derived in optical bands are translated into the extinction
laws applied in Sect. \ref{subsubsec:mean-relation} \label{tab:ew}}
\begin{tabular}{l c c c}
\hline\hline 
Works & $\rm \EJKs/EW$         & std. deviation \\ 
      & $\rm (mag\,\AA^{-1})$  &                \\ [0.5ex]
\hline 
This work\tablefootmark{a} & 1.874 & 0.15 \\ 
This work\tablefootmark{b} & 1.802 & 0.26 \\
This work\tablefootmark{c} & 1.884 & 0.22 \\
\citet{Sanner78}           & 1.964 & 0.11 \\ 
\citet{Munari00}           & 1.853 & 0.03 \\
\citet{Wallerstein07}      & 3.176 & 0.56 \\
\citet{Munari08}           & 1.874 & 0.03 \\
\citet{Kos13}              & 1.716 & 0.23 \\
\citet{Puspitarini15}      & 1.461 & -- \\ [0.5ex]
\hline
\end{tabular}
\end{center}
\tablefoot{ \\
\tablefoottext{a}{reddening bins} \\
\tablefoottext{b}{individual fields} \\
\tablefoottext{c}{after correction}
}
\end{table}

\begin{figure*}
  \centering
  \includegraphics[width=8.4cm]{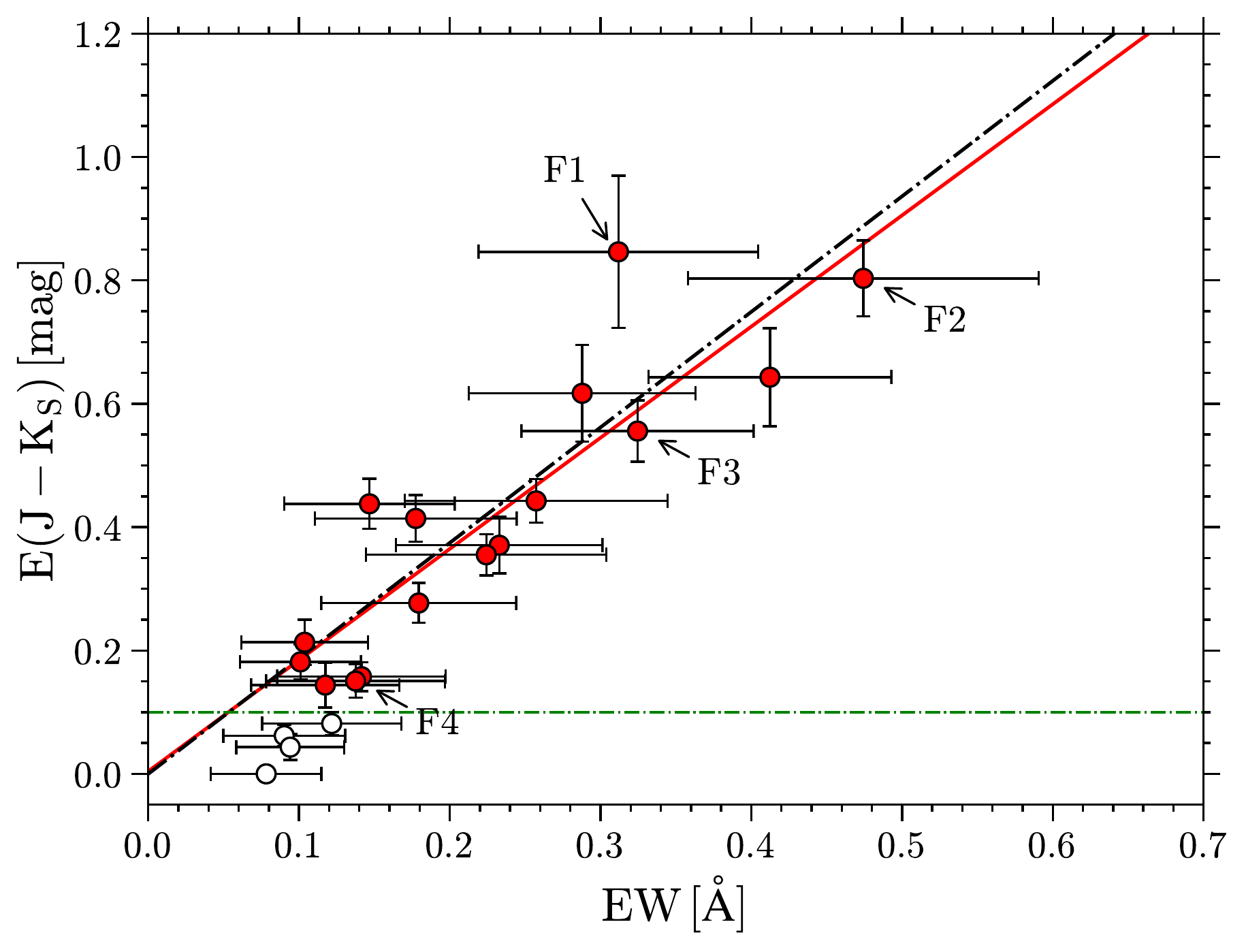}
  \includegraphics[width=8.4cm]{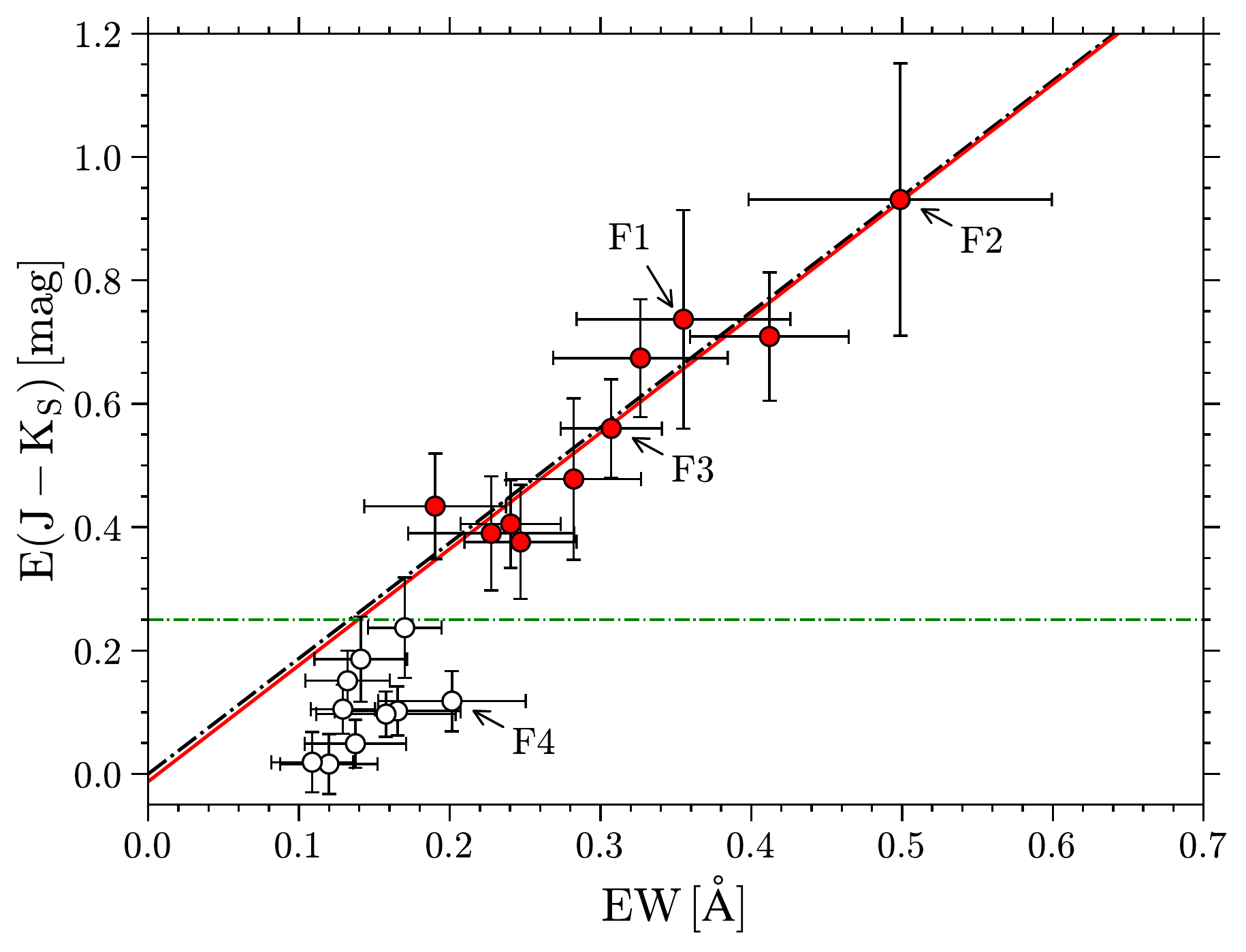}
  \caption{Correlation between the EW of the {\it Gaia DIB} and the reddening $\EJKs$ derived from individual GIBS
  fields before (left panel) and after (right panel) the correction. The red lines are fit to the
  red dots in each panel, and white dots are discarded. The dot-dashed black lines present the relation
  derived by \citet{Munari08}. The dot-dashed green lines indicate $\EJKs\,{=}\,0.10$ and 0.25\,mag in left and 
  right panels, respectively. Four fields are indicated in each panel for comparison, and a detailed discussion
  is presented in Section \ref{subsubsec:ini-field} and Appendix \ref{appsec:ext-cali}.}
  \label{fig:ew-field}
\end{figure*}

\begin{figure}
  \centering
  \includegraphics[width=8.4cm]{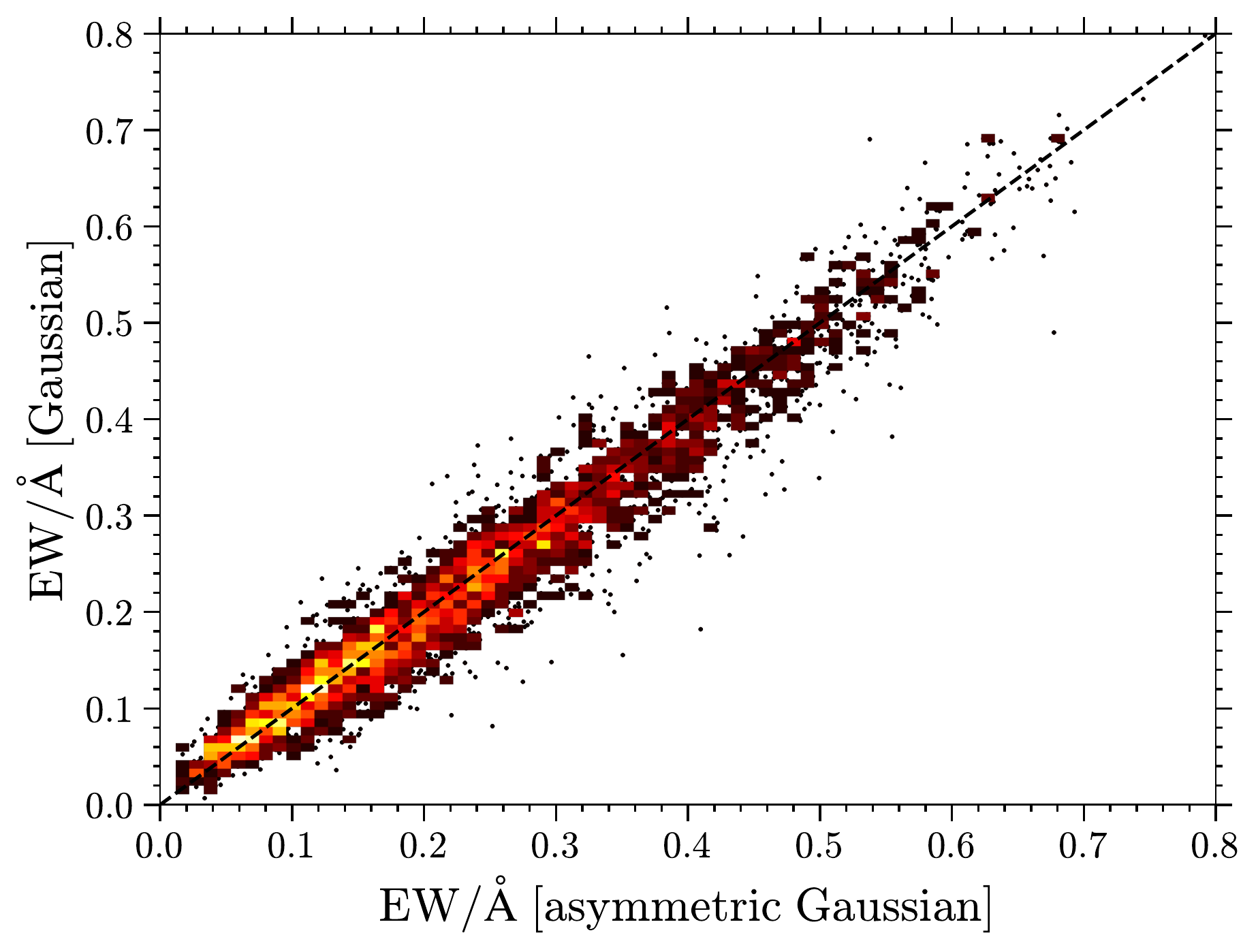}
  \caption{Comparison between the EWs of Gaussian and asymmetric Gaussian fits. The color represents
  the number density. The dashed black line traces the one-to-one correspondence.}
  \label{fig:asym}
\end{figure}

\begin{figure}
  \centering
  \includegraphics[width=8.4cm]{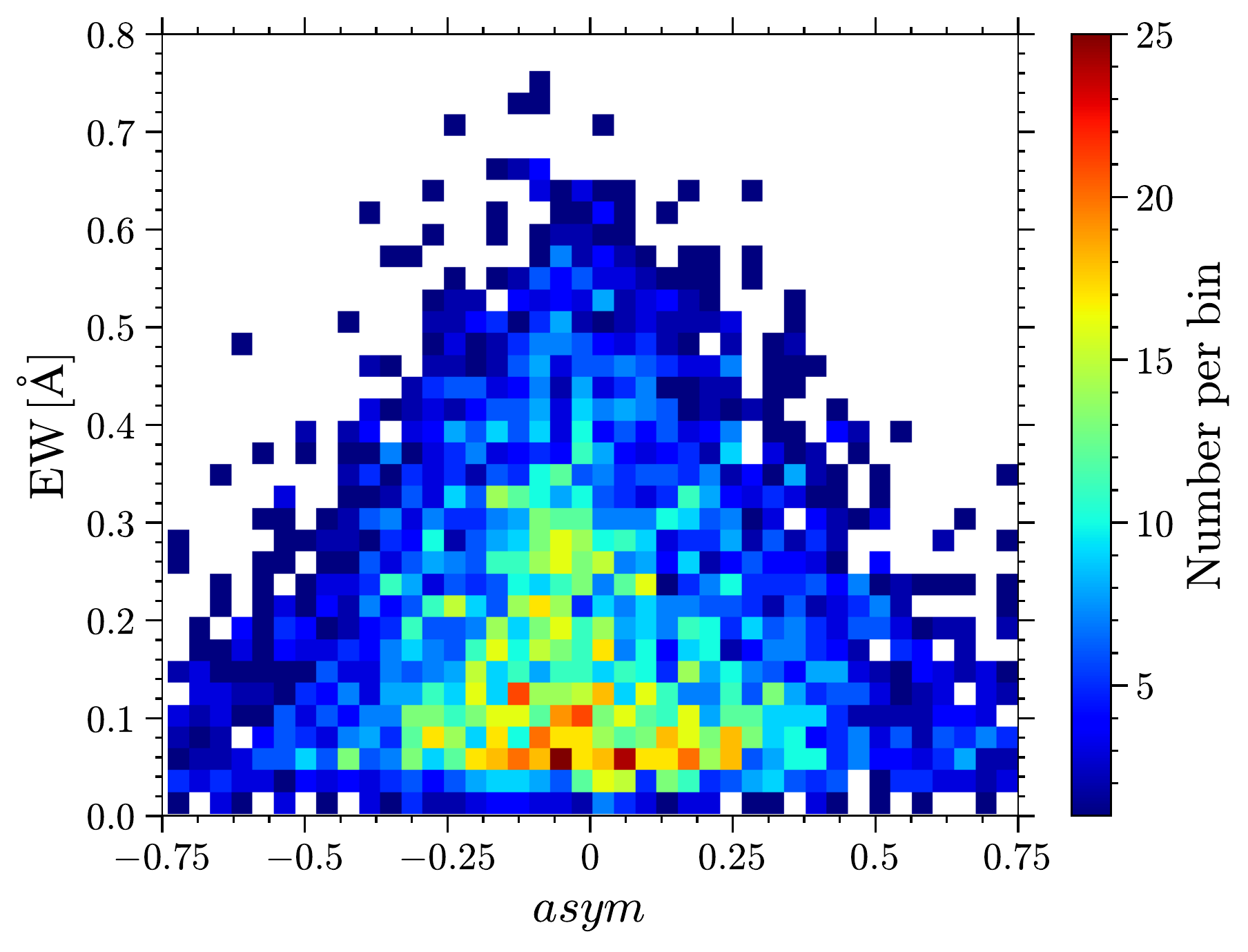}
  \caption{EW as a function of the asymmetry ($asym$). The color shows the number of stars in
  each calculated bin.}
  \label{fig:asym-ew}
\end{figure}

\begin{figure*}
  \centering
  \includegraphics[width=16.8cm]{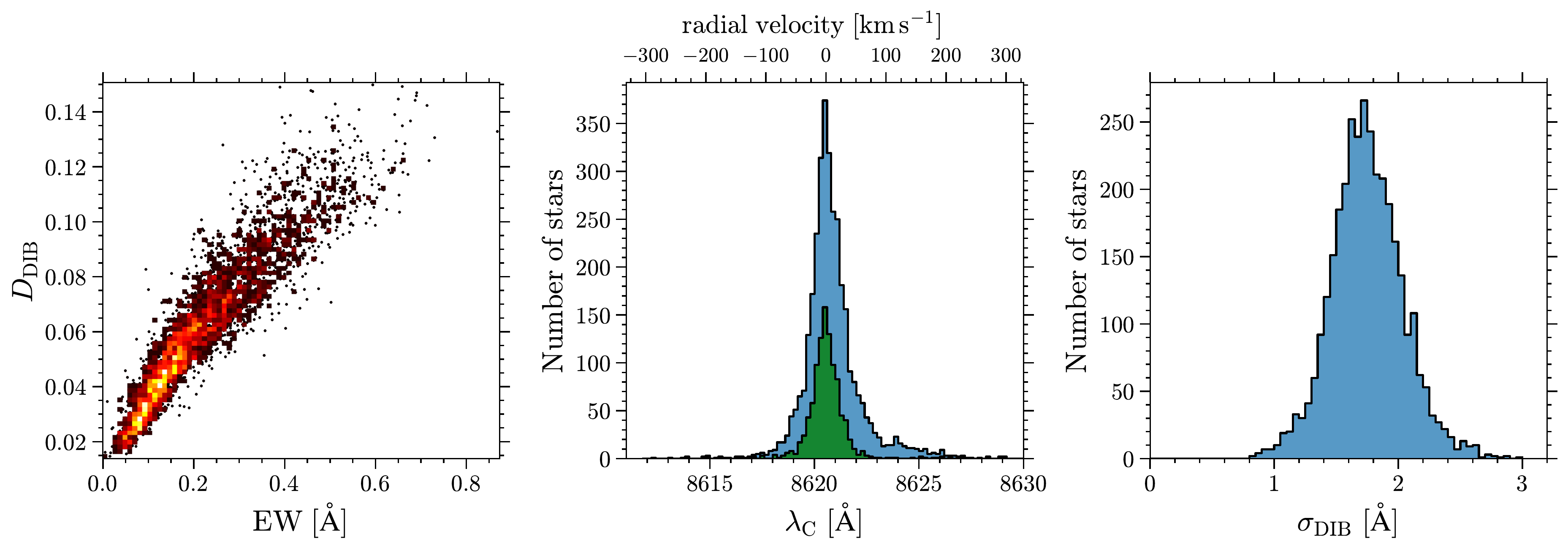}
  \caption{Distributions of the DIB parameters. The left panel shows the depth $D$ vs. EW. The color represents
  the number density. The middle panel is the measured line center $\lambda_C$ in the heliocentric frame. The upper axis shows the 
  corresponding radial velocities of the DIB carriers. The green histograms present the subsample selected to 
  determine the rest-frame wavelength (see Sect. \ref{subsubsec:rest-wave}). The right panel presents the histograms
  of the width $\sigma$.}
  \label{fig:dib-shape}
\end{figure*}

\subsubsection{EW versus $\EJKs$ relation from individual GIBS fields} \label{subsubsec:ini-field}

We decreased the EW dispersion by averaging the measurements by calculating the median values of 
the EW and $\EJKs$ for each GIBS field individually. The correlation between EW and $\EJKs$ derived from the 20 GIBS 
fields is shown in Fig. \ref{fig:ew-field} (left panel). After discarding fields with $\EJKs\,{<}\,0.1$ 
(indicated by the dot-dashed green line), we gain a linear relation of  
$\EJKs\,{=}\,1.802\,({\pm}\,0.258)\,{\times}\,{\rm EW}\,{+}\,0.004\,({\pm}\,0.065)$,
corresponding to $\frac{\EBV}{\rm EW}\,{=}\,2.615$, which is slightly smaller than the value derived before (2.721).
The EW dispersion in each field does not notably decrease compared to that in the 
reddening bins. A possible reason is that the low spatial resolution of $\EJKs$ obscures 
the environmental variation in each GIBS field, which is traced by the {\it Gaia DIB}, leading to large 
dispersion of EW but small dispersion of $\EJKs$ (the dispersion of $\EJKs$ is usually smaller than 
its uncertainty of individual targets). This cannot account for the large dispersion in the
fields with large EW, however, especially for the field $(l,b)\,{=}\,(8,-2)$ (indicated by ``F1'' in Fig. \ref{fig:ew-field}). 

The problem may come from the contamination of stars in the GIBS target selection that are not RC stars.
The RC stars in the GIBS survey are selected based on the $J~\text{versus }~(J-K_{\rm S})$ color-magnitude diagram (CMD)
with a limit in the $J$ magnitude and a lower cut of $J-K_{\rm S}$, while 
$J-K_{\rm S}$ is not stringently constrained at the red end \citep[see Fig. 3 in][]{Zoccali14}.
Therefore the RC sample might be contaminated by highly reddened dwarfs and/or RGB stars. 
The spectra of the contaminators that might be very different from the RC template may
give rise to pseudo-features on the interstellar spectra, causing incorrect fittings and calculations of EW.
Although $\EJKs$ we used for the targets come from S20, which is not sensitive to stellar type, 
their map was calibrated by \citet{Gonzalez12}, which was based on RC stars. This means that contaminators far away from 
the peak color $(J-\Ks)$ will also gain incorrect $E(J-\Ks)$ values.

We therefore performed the correction for both reddening and EW based on VVV--DR2 catalog\footnote{We accessed the data by 
SIMBAD/VizieR-II/348, https://vizier.u-strasbg.fr/viz-bin/VizieR-3?-source=II/348/vvv2} \citep{Minniti17}. 
We constructed a purer sample of RC stars by applying an additional color cut. For each field, RC candidates 
were first selected from the VVV catalog in a circular region located at the field center with a radius of $0.5\deg$. 
Then we fit the $(J-\Ks)$ colors within the range of the $J$ magnitudes given by the GIBS targets (see the dashed orange
lines in Fig. \ref{fig:ext-cali}) with a Gaussian function to obtain the peak color as well as the $1\sigma$ width.
This criterion ensures that our RC sample is as pure as possible, with the disadvantage that we loose stars.
The percentage of the rejected stars differs for different fields. In total, we obtained 2437 targets in the purer sample, 
compared to 4194 in original sample. This means that about 42\% stars are discarded. Assuming an intrinsic color of the RC stars of 
$(J-K_{\rm S})_0\,{=}\,0.674$ \citep{Gonzalez11}, we obtain an average $\EJKs$  for each GIBS field where the 
standard deviation of $J-K_{\rm S}$ is treated as the uncertainty of $\EJKs$, $\Delta \EJKs$, including not only 
the error of $\EJKs$ calculated by RC stars, but also the dispersion of $\EJKs$ in each field. For low-reddening fields, 
$\Delta \EJKs$ are similar to the mean errors of $\EJKs$ given by S20. The average value of $\Delta \EJKs$, 0.066, is also
close to the resultant error of RC stars when a photometric error of 0.03 for the $J$ and $\Ks$ bands and a spread of
$(J-\Ks)_0$ of 0.03 is assumed. The increase in $\Delta \EJKs$ in highly extincted fields is caused by the $J$--magnitude range
applied for GIBS targets. A  wider $J$--magnitude range covers a wider range of extinction and results in a larger dispersion 
of $\EJKs$. $\Delta \EJKs$ is consequently dominated by the dispersion and can reach values above 0.1\,mag.

Furthermore, the median EW of each field was recalculated by only considering the targets with $J-\Ks$ within the $1\sigma$ region 
(dashed green lines in Fig. \ref{fig:ext-cali}). The correlation between EW and $\EJKs$ for individual fields after 
correction is presented in the right panel in Fig. \ref{fig:ew-field}. The dispersion of EW in each field markedly decreases 
and now is comparable to the uncertainty of $\EJKs$. A tighter linearity of the correlation was also derived by considering 
fields with $\EJKs\,{>}\,0.25$ with a Pearson correlation coefficient of 0.95, compared to the value 0.88 before the correction.
The coefficient of the linear relation is $\EJKs\,{=}\,1.884\,({\pm}\,0.225)\,{\times}\,{\rm EW}\,{-}\,0.012\,({\pm}\,0.072)$, 
corresponding to $\frac{\EBV}{\rm EW}\,{=}\,2.734$, which is highly consistent with \citet{Munari08} as well. Using our new purer sample 
(see Fig. \ref{fig:purer-RC}), we obtain a similar relation (as discussed in detail in Appendix \ref{appsec:ext-cali}). The relative strength 
($\rm EW/\Av\,{\approx}\,0.1$\,{\AA}\,$\rm mag^{-1}$) and the tight correlation with reddening confirm that {\it Gaia DIB} is a powerful 
tracer of ISM species, independent of the foreground extinction, as suggested by \citet{Kos13}.

Four fields were selected to demonstrate the correction effect: F1 $(l,b)\,{=}\,(8,-2)$, F2 $(l,b)\,{=}\,(0,-1)$, F3 $(l,b)\,{=}\,(-5,-2)$,
and F4 $(l,b)\,{=}\,(-4,-6)$. They are indicated in Fig. \ref{fig:ew-field}. The applied correction performs 
very well for highly extincted regions. However, it intensifies the deviation of the fields with low extinctions because 
a) the S/N of the GIBS target limits the detection of small DIBS (see also Fig. \ref{fig:simu1}), and
b) at high latitudes where extinction is lower, $J-K_{\rm S}$ is not very sensitive to small-scale 
variation in the reddening. Optical data such as $B-V$ would be more sensitive.

\subsection{Properties of the {\it Gaia DIB}} \label{subsec:dib-property}

\subsubsection{Asymmetry of the DIB profile} \label{subsubsec:dib-asym}

Different DIBs have diverse profiles, from single profiles \citep{Sarre95} to resolved substructures 
\citep{Galazutdinov02}. Asymmetric shapes originate from the distortion of unresolved substructures 
or blended DIBs \citep{Kos17}. The {\it Gaia DIB} is suggested to be blended by two DIBs \citep{Jenniskens94}, 
while no intrinsic asymmetry has been unveiled by previous works. For most of the GIBS results, the difference 
between Gaussian and asymmetric Gaussian EW is comparable to their uncertainties (Fig. \ref{fig:asym}). 
As shown in Fig. \ref{fig:asym-ew}, large asymmetries only occur in small DIBs, which probably
originate in noise. No signature of intrinsic asymmetry of the {\it Gaia DIB} is revealed by Figs. \ref{fig:asym}
and \ref{fig:asym-ew} for large or small EWs.

\subsubsection{Rest-frame wavelength} \label{subsubsec:rest-wave}

The rest-frame wavelength of the {\it Gaia DIB} is reported as 8620.8\,{\AA} by \citet{Galazutdinov00} 
from one single star, and 8621.2\,{\AA} by \citet{Jenniskens94} from four hot stars, and  8620.4\,{\AA}
by \citet{Munari00} and \citet{Munari08} from dozens of RAVE hot stars. The determination of 
\citet{Munari08} was based on the assumption that the average velocity of their carriers, which are 
close to the Galactic center, is essentially zero, after adopting the ISM radial velocity map of 
\citet{BB93}. Therefore the average central wavelength $\lambda_C$ represents the rest-frame
wavelength.

We followed this method and selected spectra from the GIBS fields with $-3^{\circ}\,{<}\,b\,{<}\,3^{\circ}$ and 
$-6^{\circ}\,{<}\,l\,{<}\,3^{\circ}$. Finally, we obtained 1015 spectra.
The observed $\lambda_C$ is in the stellar frame, and it 
has to be converted into the heliocentric frame based on the stellar radial velocity.
The converted $\lambda_C$ presents a Gaussian distribution (Fig. \ref{fig:dib-shape}, middle
panel, green histograms) with a mean value of 8620.55\,{\AA} and a standard deviation of 0.55\,{\AA}. 
Our derived rest-frame wavelength is close to the result of \citet{Munari08}, but the large 
uncertainty makes it still not sufficiently definite. A more accurate method is investigating 
measurements toward the Galactic anticenter, as illustrated in \citet{Zasowski15}, which we 
intend to apply to the Gaia--RVS spectra.

\subsubsection{Parameter distributions} \label{subsubsec:dib-para}

Figure \ref{fig:dib-shape} shows the distributions of the measured DIB parameters from the GIBS spectra: depth $D$ 
versus EW, line center $\lambda_C$, and width $\sigma$. Unreasonable results with very small or large measurements were 
eliminated. Small EWs increase with depth, while the relation deviates from linearity 
when $\rm EW>0.2$\,{\AA}. Profiles with $D\,{>}\,0.15$ come from spectra without proper normalization and were discarded. 
The limit of $D$ in the QF test (Sect. \ref{subsec:QF}) was therefore set as 0.15. 

The peak value of $\lambda_C$ for all the GIBS targets is the same as that of the subsample for deriving the rest-frame
wavelength. The distribution of the subsample can be well fit by a Gaussian function. However, the $\lambda_C$ distribution 
of all the GIBS targets apparently deviates from the Gaussian profile 
and contains a bump around 8624\,{\AA}. The origin of this second bump is not clear and will be studied in a 
forthcoming paper.

The valid widths are within 0.6--3.2\,{\AA}, with a peak value of 1.74\,{\AA}. This value is close to the
peak value of the APOGEE DIB $\lambda$1.5273 derived by \citet{Zasowski15}. These two DIBs also have a similar
relative strength. An abrupt decrease in $\sigma$ occurs at $\sim$1.4\,{\AA}; it is larger than that of the APOGEE DIB
but consistent with APOGEE measurements for sightlines with $l\,{<}\,40\deg$. As the inner Galaxy generally contains 
broader features than the outer Galaxy \citep{Zasowski15}, we set a lower value $\sigma\,{=}\,1.2$ for the QF test
to distinguish narrow and shallow DIBs. We did not find any DIB width larger than 3.2\,{\AA} but a steep
decrease from 2 to 3. However, \citet{Zasowski15} revealed that the APOGEE DIB can be as broad as 6\,{\AA} 
toward the Galactic center. The upper limit of $\sigma$ in the QF test could change if we were to find physical DIB
profiles with larger width in further studies.

\section{Conclusions} \label{sec:conclusion}

The main goal of this work was to develop a procedure for the automatic detection and measurement of the {\it Gaia DIB} 
($\lambda\,{\approx}\,8620$\,{\AA}). 
A preliminary detection was applied to
exclude low-S/N spectra and/or DIBs below the detection limit. The DIB feature was extracted from the cool-star spectra
using  synthetic spectra, while for hot stars, we applied a specific
model based on GP \citep{Kos17} to directly measure the DIB feature on the observed spectra without any stellar
templates. The DIB profile was fit by a Gaussian function, and the EW was also calculated. A simulation based on pseudo-spectra 
with different S/N illustrated the effect of the random noise on the DIB fitting, as well as the EW calculation. 
Based on these simulations, a minimum S/N of 50 is required to detect DIBs. The error contributed by the synthetic spectra 
for cool stars and local continua for hot stars was also considered through a second fit. Furthermore, some tests on fitted 
parameters, $\{D,\lambda_C,\sigma\}$, similar to \citet{Elyajouri16}, were used to assess their qualities. 

These procedures and techniques were applied on a sample of 4979 GIBS spectra. The main results are summarized below.

By taking the median values from different reddening bins, we derived a linear relation between the EW of the 
{\it Gaia DIB} and the reddening: $\EJKs\,{=}\,1.875\,({\pm}\,0.152)\,{\times}\,{\rm EW}\,{-}\,0.011\,({\pm}\,0.048)$. Applying 
the CCM89 model and the NIR extinction laws toward Galactic center from \citet{Nishiyama09}, we find $\EBV/{\rm EW}\,{=}\,2.721$,
which is highly consistent with the results of \citet{Munari00} (2.69) and \citet{Munari08} (2.72). Additionally, the 
difference of the coefficient between our result and other studies with different sightlines implies a possible 
variation in the relation for different ISM conditions.

The median measurements from individual GIBS fields presented a relation with a coefficient of $\EJKs/{\rm EW}\,{=}\,1.802\,{\pm}\,0.258$, with 
a relatively large dispersion of the EW in each field due to the contamination of non-RC stars in the GIBS sample. We 
eliminated them by using an additional color cut. This led to a smaller dispersion and improved the linearity of 
the EW--$\EJKs$ correlation for individual fields. The corrected relation, $\EJKs\,{=}\,1.884\,({\pm}\,0.225)\,{\times}\,{\rm EW}\,{-}\,0.012\,({\pm}\,0.072)$, 
also compares well with other results.

Assuming that the average radial velocity of the DIB carrier is zero when they are distributed close to the Galactic center 
\citep{BB93}, we determined the rest-frame wavelength of the {\it Gaia DIB} as $\lambda_0\,{=}\,8620.55 \pm 0.55$\,{\AA}. 

We also fit the GIBS spectra with an asymmetric Gaussian model. The results are in general consistent with those 
from the Gaussian model. No intrinsic asymmetry is found. This shows that the Gaussian profile is a proper assumption for 
the {\it Gaia DIB} and can be applied to the spectra from other spectroscopic surveys.

\begin{acknowledgements}
We thank the anonymous referee for very helpful suggestions and constructive comments.
We would like to thank Dr. Janez Kos for his help with the Gaussian process technique and his useful suggestions.
This work made use of the data from the sky surveys of GIBS and VVV. Funding for RAVE has been provided by: the Leibniz Institute for 
Astrophysics Potsdam (AIP); the 
Australian Astronomical Observatory; the Australian National University; the Australian Research Council; the French National 
Research Agency; the German Research Foundation (SPP 1177 and SFB 881); the European Research Council (ERC-StG 240271 Galactica); 
the Istituto Nazionale di Astrofisica at Padova; The Johns Hopkins University; the National Science Foundation of the USA (AST-0908326); 
the W. M. Keck foundation; the Macquarie University; the Netherlands Research School for Astronomy; the Natural Sciences and Engineering 
Research Council of Canada; the Slovenian Research Agency; the Swiss National Science Foundation; the Science {\&} Technology 
FacilitiesCouncil of the UK; Opticon; Strasbourg Observatory; and the Universities of Basel, Groningen, Heidelberg and Sydney. 
EV acknowledges the Excellence Cluster ORIGINS Funded by the Deutsche Forschungsgemeinschaft (DFG, German Research Foundation) under 
Germany's Excellence Strategy \-- EXC \-- 2094 \--390783311. HZ is funded by the China Scholarship Council (No.201806040200).
\end{acknowledgements}

\bibliographystyle{aa}
\bibliography{references.bib}

\appendix

\section{Correction for individual fields} \label{appsec:ext-cali}

As a standard candle \citep{PS98}, RC stars are widely used to trace interstellar extinction \citep{Indebetouw05,GaoJ09,WC19}
and nebular distance \citep{Guver10,ShanSS18,WangS20,ChenBQ20}. RC candidates are usually selected from $(J\,{-}\,\Ks,\,\Ks)$ CMD 
with empirical borders defined by naked eyes \citep[e.g.,][]{WangS20,ChenBQ20}. To exclude the contamination of other types of stars, 
peak colors are determined in different $\Ks$ bins to form the track of RC toward a given line of sight. With a similar
method, we corrected for $\EJKs$ and EW for individual GIBS fields (Sect. \ref{subsubsec:ini-field}).

\begin{figure*}
  \centering
  \includegraphics[width=16.8cm]{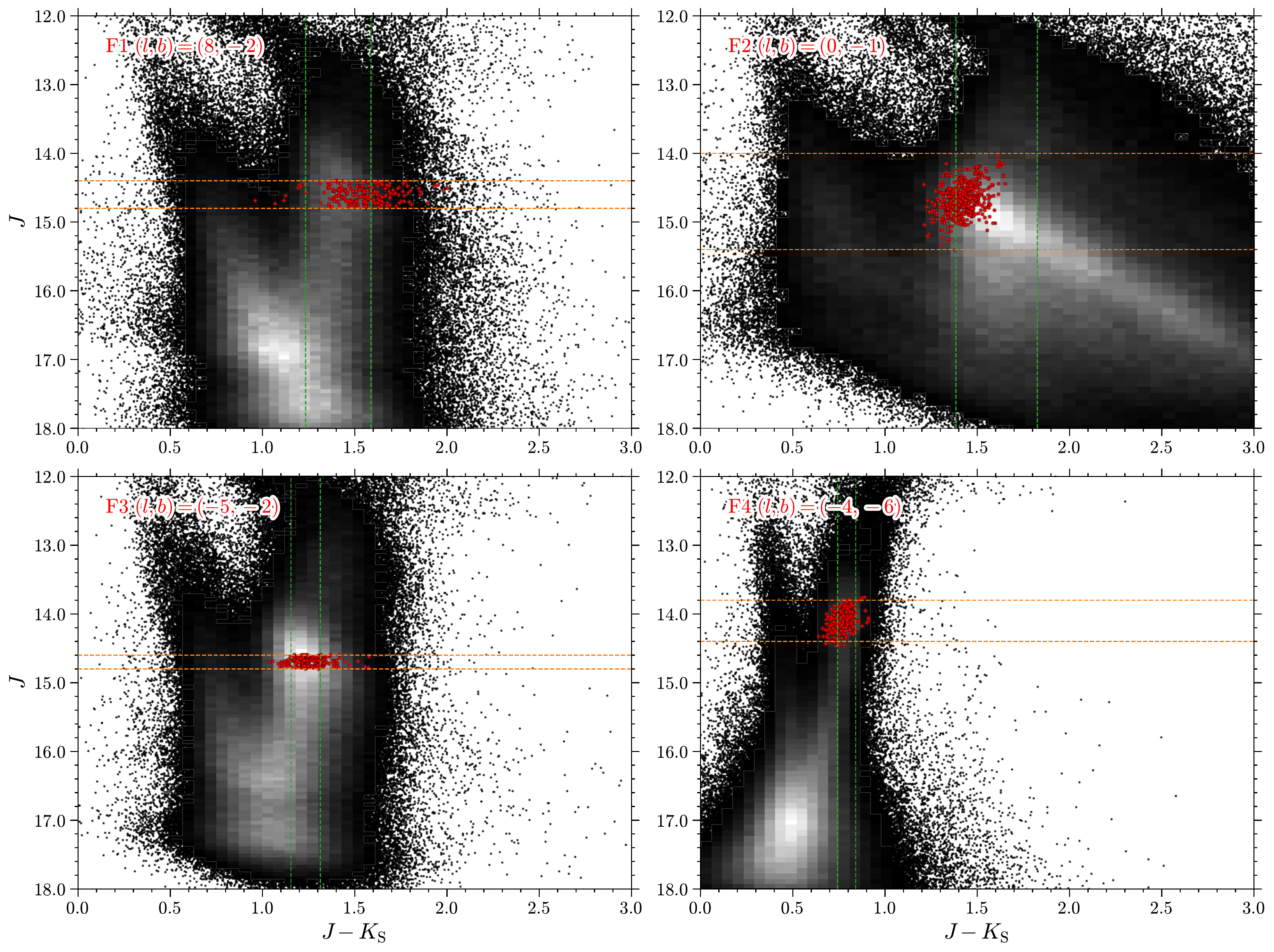}
  \caption{$(J-K_{\rm S},\,J)$ CMD for four selected GIBS fields. The gray-scale image shows the
  number density of the stars selected from VVV--DR2 catalog. The red dots are the GIBS targets in each
  field. $J\Ks$ photometry of GIBS targets are provided by the PSF photometry catalog developed by \citet{Surot19}.
  The dashed green lines indicate the $1\sigma$ extent widths around the RC peak densities for 
  the regions defined by the range of $J$ magnitudes of GIBS targets in each field (dashed orange lines).
  The name and coordinates of each field are indicated at the upper left corner in each panel.}
  \label{fig:ext-cali}
\end{figure*}

Figure \ref{fig:ext-cali} presents the $(J\,{-}\,K_{\rm S},\,J)$ CMD for four selected fields (see Sect. \ref{subsubsec:ini-field} and 
Fig. \ref{fig:ew-field}) to illustrate the tightening caused by the correction in the linearity correlation between EW and $\EJKs.$  F1 shows that 
many stars are found to be to the right of the peak color out of the $1\sigma$ region, which might be highly reddened non-RC stars, 
leading to a much higher average $\EJKs$ of F1, as shown in Fig. \ref{fig:ew-field}. The corrected reddening becomes significantly 
lower and fits the linear relation much better with the corrected EW. On the other hand, F2 contains some foreground stars (possible dwarf stars) 
located at the blue side of the peak color. After correction, $\EJKs$ and EW slightly increase and are highly consistent 
with the linear relation derived by \citet{Munari08}. The largely increased uncertainty of $\EJKs$ is due to the wide range of
$J$ magnitudes (${\approx}\,1.2$\,mag). A trend can be found toward the lower right of the peak color (the brightest part in F2 CMD in Fig. 
\ref{fig:ext-cali}). This means that the RC stars, which all lie at the bulge distance, spread in the CMD following the reddening vector 
because of the differential reddening over the selected region. Therefore the dispersion of EW in F2 is not only caused by the uncertainties, 
but implies the trace of environmental variation, which is hardly unveiled by the 2D extinction maps.

F3 have contaminators on both sides. The corrected $\EJKs$ slightly increases, but the EW is nearly unchanged, while the dispersion of EW
becomes apparently smaller after the contaminators are eliminated. This is the case in nearly all the fields. The correction can alleviate the EW dispersion.
The correction also causes the increase in the median EW for the fields with $\EJKs\,{<}\,0.2$\,mag.  F4 is typical of these fields. Its reddening decreases
by about 0.05\,mag, while its EW increases by about 0.05\,mag, so that it becomes more like an outlier. The reasons for the overestimation of EW could be that
1) small DIBs cannot be detected because of the low spectral S/N, thus the median values increase, 2) the reduction in RC density at high latitudes
makes it harder to accurately estimate the reddening for these fields, and 3) dwarfs with high reddening could have $J-\Ks$ very close to the
peak colors and can hardly be excluded. Consequently, some of the detected DIBs may be pseudo-features caused by the mismatches between
the spectra of the dwarfs and the template for RC stars. Their EWs increase the median EW of the field. This means that our correction does not
perform well for low-reddening fields at high latitudes.

\begin{figure}
    \centering
    \includegraphics[width=8.4cm]{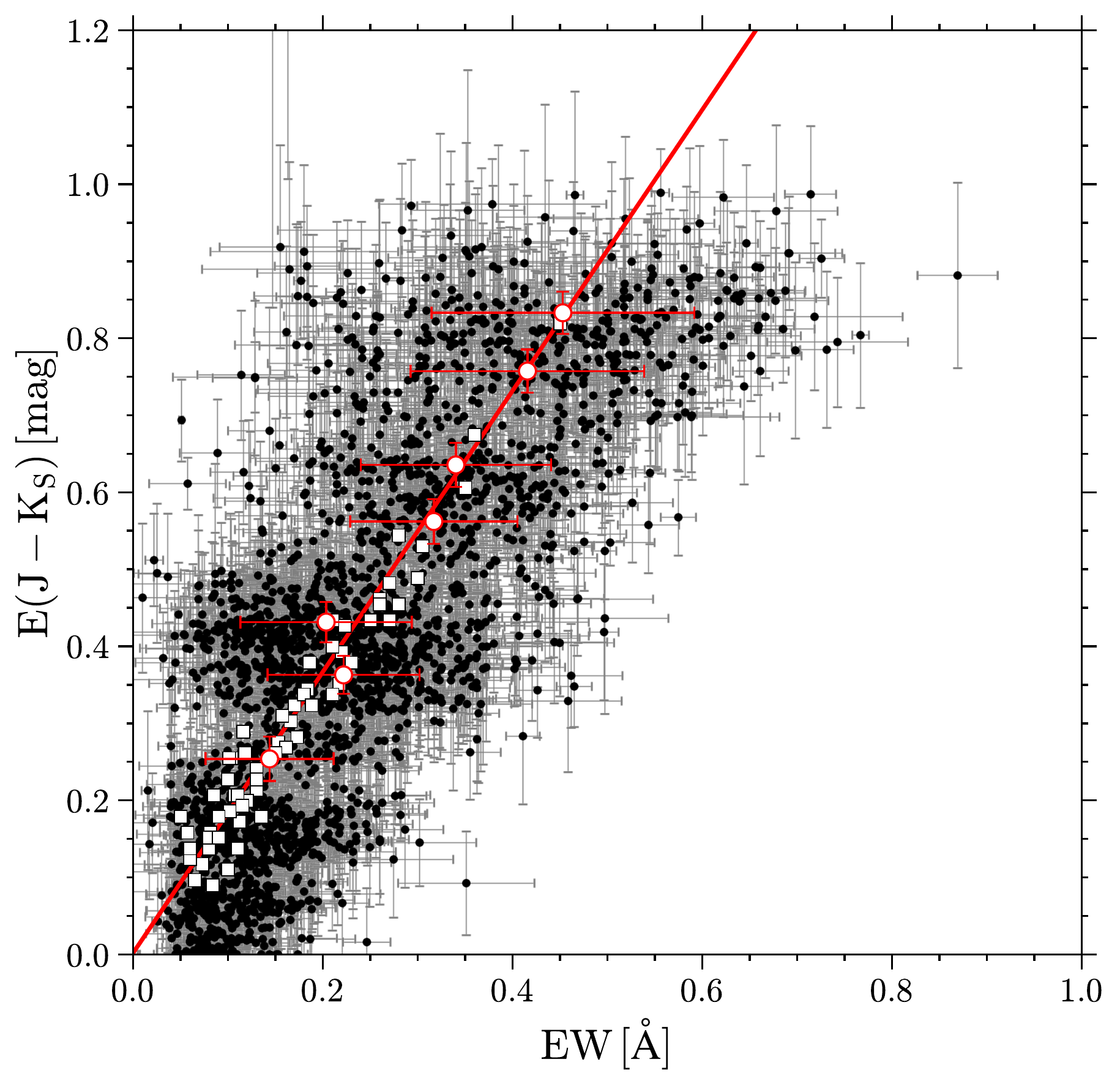}
    \caption{Same as Fig. \ref{fig:ew-mean}, but with a purer RC sample.}
    \label{fig:purer-RC}
\end{figure}

We derived the purer RC sample with targets in each field within $1\sigma$ width of $(J\,{-}\,\Ks)$. Finally, we obtained 2437 cases. The EW--$\EJKs$
relation derived by the purer RC sample is shown in Fig. \ref{fig:purer-RC}. We still derive median values of EW and $\EJKs$ with $\EJKs$ within
0.2--0.9\,mag with a step of 0.1\,mag. The linear relation is $\EJKs\,{=}\,1.825\,({\pm}\,0.123)\,{\times}\,{\rm EW}\,{+}\,0.002\,({\pm}\,0.039)$.
The coefficient is consistent (although slightly smaller)
with previous derived values, and the intercept is much closer to zero. We can also find that 1) almost no cases with $\EJKs\,{>}\,1.0$. High reddening,
especially in F1, is excluded by the correction, and 2) large EW still exist, implying that they are true features. They trace the variation in the
ISM, while $\EJKs$ failed because of the low resolution of the extinction map. 

\clearpage

\end{CJK*}

\end{document}